\documentclass{article}

\usepackage{arxiv}

\usepackage[utf8]{inputenc} 
\usepackage[T1]{fontenc}    
\usepackage{hyperref}       
\usepackage{url}            
\usepackage{booktabs}       
\usepackage{amsfonts}       
\usepackage{nicefrac}       
\usepackage{microtype}      
\usepackage{lipsum}		
\usepackage{graphicx}

\usepackage{subfigure}
\usepackage{multirow}
\usepackage{mathtools}
\usepackage{stmaryrd} 

\title{Optimize What You Evaluate With: A Simple Yet Effective Framework For Direct Optimization Of IR Metrics}

\author{ \href{https://orcid.org/0000-0002-1569-8507}{\includegraphics[scale=0.06]{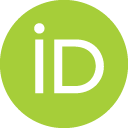}\hspace{1mm}Hai-Tao Yu}\\
	Faculty of Library, Information and Media Science\\
	University of Tsukuba\\
	1-2 Kasuga, Tsukuba City, Ibaraki, 305-8550, Japan\\
	\texttt{yuhaitao@slis.tsukuba.ac.jp}
}

\renewcommand{\shorttitle}{\textit{arXiv} Template}

\hypersetup{
pdftitle={A template for the arxiv style},
pdfsubject={q-bio.NC, q-bio.QM},
pdfauthor={David S.~Hippocampus, Elias D.~Striatum},
pdfkeywords={First keyword, Second keyword, More},
}

\begin{document}
\maketitle

\begin{abstract}
Learning-to-rank has been intensively studied and has shown significantly increasing values in a wide range of domains, such as \textit{web search}, \textit{recommender systems}, \textit{dialogue systems}, \textit{machine translation}, and even \textit{computational biology}, to name a few. The performance of learning-to-rank methods is commonly evaluated using rank-sensitive metrics, such as \textit{average precision} (AP) and \textit{normalized discounted cumulative gain} (nDCG). Unfortunately, how to effectively optimize rank-sensitive objectives is far from being resolved, which has been an open problem since the dawn of learning-to-rank over a decade ago. In this paper, we introduce a simple yet effective framework for directly optimizing information retrieval (IR) metrics. Specifically, we propose a novel \textit{twin-sigmoid} function for deriving the \textit{exact rank positions} of documents during the optimization process \textit{instead of using approximated rank positions or relying on the traditional sorting algorithms} (e.g., \textit{Quicksort} \cite{QuickSort}). Thanks to this, the rank positions are differentiable, enabling us to reformulate \textit{the widely used IR metrics} as differentiable ones and directly optimize them based on neural networks. Furthermore, by carrying out an in-depth analysis of the gradients, we pinpoint the potential limitations inherent with direct optimization of IR metrics based on the vanilla sigmoid. To break the limitations, we propose different strategies by explicitly modifying the gradient computation. To validate the effectiveness of the proposed framework for direct optimization of IR metrics, we conduct a series of experiments on the widely used benchmark collection \textit{MSLRWEB30K}. The experimental results demonstrate that: (1) Direct metric optimization is a more appropriate choice than the commonly used surrogate loss functions, such as ListMLE, ListNet and WassRank. (2) Regarding direct metric optimization, the proposed methods significantly outperform the baseline approach ApproxNDCG. Compared with the state-of-the-art tree-based approach LambdaMART, the proposed methods building upon a simple feed-forward neural network, such as AP-type3 and nDCG-type3, can achieve comparable results.
\end{abstract}


\section{Introduction}
\label{sec:Intro}
Learning-to-rank has been intensively studied and has shown great
value in many fields, such as \textit{web search}, \textit{dialogue systems}, and \textit{computational biology}  \cite{l2rbio}. In this paper, we focus on the field of \textit{document retrieval}. Following the \textit{Cranfield} experimental paradigm, a large number of queries are provided. Each query is associated with a set of documents to be ranked, of which the standard relevance
labels are also included. Each query-document pair is represented
through a feature vector. The desired scoring model (or function) assigns a score to each document, then a ranked list of documents can be obtained by sorting the documents
in descending order of scores. In general, the document with the highest
score is assigned a rank of 1. In other words, the rank position of
a document represents its relevance with respect to the query. The metrics, such as AP and nDCG \cite{FstnDCG}, are adopted to measure the performance.

The information retrieval (IR) community has experienced a flourishing
development of learning-to-rank methods, such as \textit{pointwise} methods, \textit{pairwise} methods and \textit{listwise} methods. The pointwise
methods \cite{SubsetRegression, ChuGaussian, ChuSVOR} transform the ranking problem into
a task of (ordinal) regression or classification on individual documents.
The idea is natural and many existing mature learning techniques on
classification and regression can be directly deployed. A major problem
is that the pointwise methods are agnostic to the relevance-based
order information among documents that are associated with the same
query. To make a step forward, the pairwise methods \cite{FreundBoosting, ShenPerceptron, RankSVMStruct}
were then proposed, which transform the ranking problem into a task
of pairwise classification. However, the loss functions merely consider
the relative order between two documents rather than the total order
relationship among all documents associated with the same query. Moreover, the number of document pairs per
query may differ from query to query, thus the result can be biased
in favor of queries with more documents in the training data.

\begin{figure}[!htbp]
	\begin{centering}
		\subfigure[The sigmoid function. \label{subfig:vanillaSigmoid}]{
			\centering{}\includegraphics[width=.32\textwidth, totalheight=1.5in]{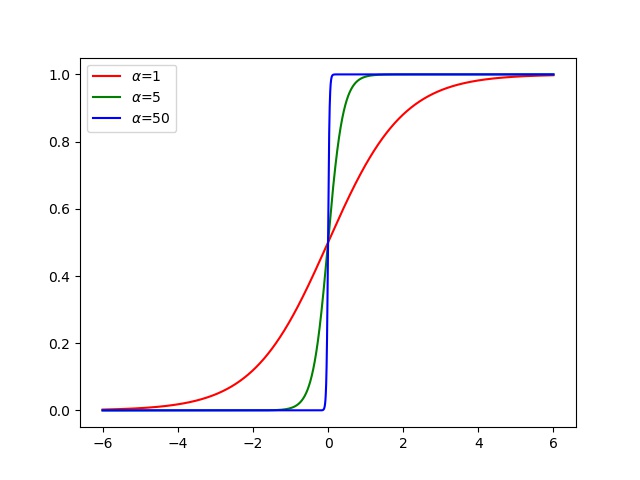}}
		\subfigure[The derivative of sigmoid. \label{subfig:derivativeOfVanillaSigmoid}]{
			\centering{}\includegraphics[width=.32\textwidth, totalheight=1.5in]{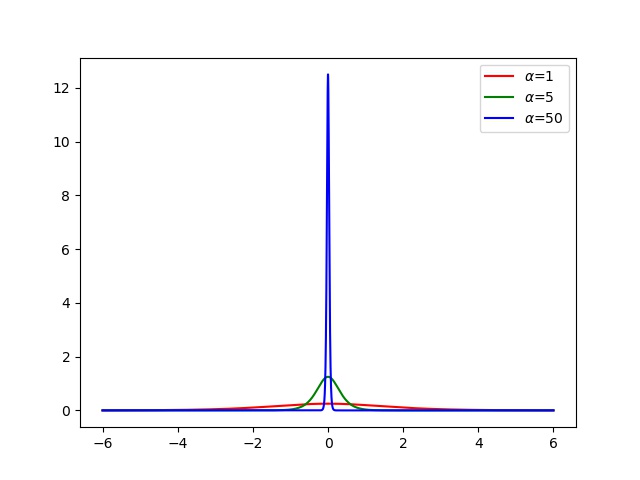}}		
		\subfigure[The twin-sigmoid function. \label{subfig:twinSigmoid}]{
			\centering{}\includegraphics[width=.32\textwidth, totalheight=1.5in]{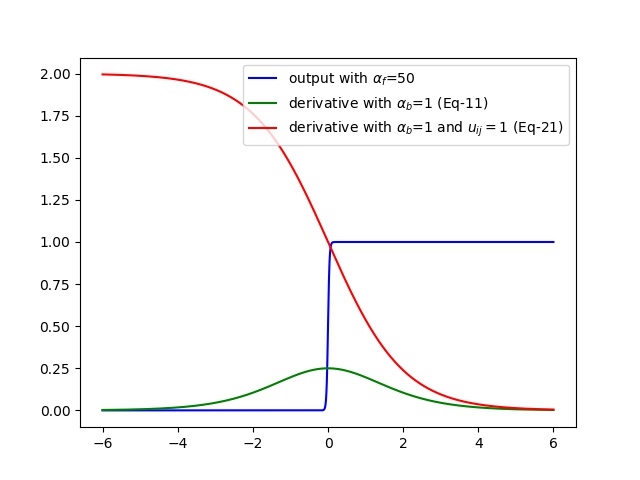}}
		\protect\caption{The plot for sigmoid and twin-sigmoid. (a) $\sigma$ determines the shape (steepness) of the sigmoidal curve. (b) For a larger $\sigma$, the derivative of sigmoid becomes quite steep. (c) For twin-sigmoid, $\sigma_f$ controls its output. Its derivative is computed in different ways, such as Eq-\ref{Eq:TwinsigmoidD1}, Eq-\ref{Eq:type2} and Eq-\ref{Eq:type3}.\label{fig:HPEffects}}
		\par\end{centering}
\end{figure}

To overcome the shortcomings of the aforementioned two categories
of ranking methods, the listwise methods \cite{OlivierMargin, AdaRank, YueSVMAP, GuiverSoftRank, SoftRank, QinApproximateNDCG, LambdaMART, ListNet, ListMLE, BoltzRank, LambdaRank} appeal to the loss function that is defined over all documents associated with the same query. The studies \cite{ListNet, ListMLE, QinApproximateNDCG, ListwiseNeuralDemo} have demonstrated that the listwise
approaches commonly show superior performance over the other
two categories of pointwise and pairwise. In this regard, the listwise methods can further be grouped into two different sub-categories. The first sub-category ignores the IR metrics during the training process. The hope is that the desired metric performance (e.g., nDCG) would in turn be maximized by minimizing some surrogate loss functions. Examples include ListMLE  \cite{ListMLE} and WassRank  \cite{TaoWSDM2019}. The second sub-category aims to directly optimize the IR metrics. This direction seems more straightforward and appealing, since what is used for performance evaluation is exactly an IR metric. As widely known, the IR metrics, such as AP and nDCG, depend on the positions at which documents are ranked. The rank information is commonly obtained via a traditional sorting algorithm (e.g., \textit{Quicksort} \cite{QuickSort}). Unfortunately, the traditional sorting algorithms are inherently undifferentiable. In particular, when we make small changes to the model parameters of a scoring function, the output scores will typically change smoothly. In contrast, the ranks of documents will not change until the documents' scores exceed one another. Hence the IR metrics will make a discontinuous change. In other words, the IR metrics are non-smooth with respect to the model parameters, being everywhere either flat (with zero gradient) or discontinuous. To overcome the aforementioned issues, many approaches have been proposed to find differentiable surrogate losses that either are loosely related to or upper-bound ranking metrics. We detail the typical approaches in Section \ref{sec:ReWork}. Despite the success achieved by the state-of-the-art methods, there are some serious limitations. First, approximating the indicator with a vanilla sigmoid \cite{QinApproximateNDCG} or softmax \cite{OlivierSmoothedIRMetric} gives rise to the following \textit{dilemma}: Take the indicator approximation based on sigmoid  \cite{QinApproximateNDCG} for example, as shown in Figure \ref{subfig:vanillaSigmoid} and Figure \ref{subfig:derivativeOfVanillaSigmoid}, as $\alpha$ becomes larger, the sigmoid approximates the indicator more closely. But the gradient becomes larger at the same time, which makes it hard to train the ranking model due to the potential issue of gradient explosion. Second, the previous methods (e.g., \cite{LMarginORM, DirOptRM, OptRM, YueSVMAP, SoftRank, GuiverSoftRank, OlivierSmoothedIRMetric, BoltzRank}) are \textit{limited to either one specific metric or two. For a new metric, the upper-bound function or the method has to be designed again}. Third, the metric scores being used as optimization objectives in previous studies \cite{OlivierMargin, AdaRank, YueSVMAP, SoftRank, GuiverSoftRank, OlivierSmoothedIRMetric, QinApproximateNDCG, SmoothHinge} are \textit{approximations rather than the real metric values}. In a nutshell, it is still an open problem on how to effectively perform direct optimization of IR metrics.

The aforementioned drawbacks motivate us to perform direct optimization of IR metrics in a novel way. Inspired by the recent work \cite{AppBridges, OptBlackboxMetric, L2LGD} on learning with \textit{detached gradients}, in this paper, we propose the novel \textit{twin-sigmoid} function, which consists of a forward component and a backward component. In particular, during the forward pass, the forward component is responsible for generating the output. During the backward pass, the backward component is responsible for generating a meaningful training signal (e.g., gradient) for the preceding layers. Armed with twin-sigmoid, we can obtain the rank positions of documents with respect to the same query given their predicted relevance scores. Thanks to this, the rank positions are differentiable, enabling us to derive the differentiable reformulations of the widely used IR metrics. The main contributions of this paper are summarized as follows:\\
\indent \textbf{(1)} Instead of relying on a traditional sorting algorithm (e.g., \textit{Quicksort} \cite{QuickSort}) or using an approximated approach, we propose a novel way to obtain the documents' rank positions based on the newly proposed \textit{twin-sigmoid} function. Furthermore, we present the differentiable reformulations of the widely used IR metrics, including precision, AP, nDCG and normalized expected reciprocal rank (nERR).\\
\indent \textbf{(2)} Regarding the stochastic optimization of rank-sensitive metrics, we carry out an in-depth analysis of the gradients, and pinpoint the potential limitations inherent with direct optimization of IR metrics based on the vanilla sigmoid. To break the limitations, we propose different methods to modify the gradients by incorporating ground-truth labels and virtual gradient.\\
\indent \textbf{(3)} To validate the effectiveness of the proposed framework for direct optimization of IR metrics, we conduct a series of experiments based on a benchmark dataset. Compared with the previous methods for direct optimization of IR metrics, the proposed framework shows better performance. Compared to the state-of-the-art approach LambdaMART, the performance of the proposed framework building upon a simple feed-forward neural network is also competitive. By discussing the pros and cons of each method, we shed new light on the nature of direction optimization of IR metrics.	

The remainder of the paper is structured as: In Section 2, we survey the prior studies on direct optimization of IR metrics and neural ranking models. In Section 3, we give the mathematical formulation of the Cranfield learning-to-rank framework. In Section 4, we present the differentiable reformulations of widely used IR metrics. In Section 5, we show different ways for stochastic optimization of rank-sensitive metrics. A series of experiments are discussed in Section 6 and Section 7. We conclude the paper in Section 8.

\section{Related Work}
\label{sec:ReWork}
In this section, we detail the related work on direct optimization of IR metrics and neural ranking models.

\textbf{Direct optimization of IR metrics} has given rise to a body of studies that attempt to find differentiable surrogate losses that either are loosely related to or upper-bound ranking metrics. The first group of methods \cite{LMarginORM, DirOptRM, OptRM, YueSVMAP, AdaRank, LambdaLossFramework} try to optimize the upper bounds of IR metrics as surrogate objective functions. For example, $SVM^{map}$ \cite{YueSVMAP} optimizes the upper bound of $1-AP$ with respect to the predicted ranking. Structured SVM is adopted to iteratively optimize the most violated constraints. The other metrics (e.g., nDCG) can be optimized in a similar way. However, the scalability is hindered since the adopted cutting plane training algorithm requires a costly identification of the most violated constraint. Xu and Li \cite{AdaRank} use an exponential loss function to upper bound $1-AP$ and $1-nDCG$. Following the boosting strategy, they repeatedly construct weak rankers by re-weighting queries and a linear combination of the weaker rankers are finally used for making predictions. Xu et al. \cite{XuDirOptMetrics} conducted a general analysis by categorizing the commonly used upper-bounds into two types. Wang et al. \cite{LambdaLossFramework} propose a
probabilistic framework for approximate metric optimization, and the Expectation-Maximization procedure is deployed to optimize metric-driven loss functions. Instead of using upper bounds or approximations, the second group of methods \cite{KDDOptRM, HEClonal} resort to different optimization techniques. For example, Tan et al. \cite{KDDOptRM} used an iterative coordinate ascent method to optimize the nDCG metric. In contrast, the third group of methods \cite{SoftRank, GuiverSoftRank, OlivierSmoothedIRMetric, QinApproximateNDCG, SmoothHinge, BoltzRank, NDCGTiedScores} use the smoothed IR metric as the optimization objective. For example, the methods \cite{BoltzRank, OlivierSmoothedIRMetric, SoftRank, HamedL2rMetric} have been developed to optimize the expectation of a target evaluation metric. Specifically, the \textit{SoftRank} method by Taylor et al. \cite{SoftRank} assumes that the relevance score of a document is governed by a Gaussian distribution, which makes it possible to derive the rank distribution of a document. Based on the rank distributions of documents associated with the same query, SoftRank optimizes the expectation of nDCG. The later work by Guiver and Snelson \cite{GuiverSoftRank} extend SoftRank by using Gaussian Process to express score uncertainties. Chapelle and Wu \cite{OlivierSmoothedIRMetric} obtain the differentiable version of nDCG based on the smoothed indicator variable $e_{ij}$, which refers to whether document $i$ is ranked at the $j$-th position. Qin et al. \cite{QinApproximateNDCG} propose a typical framework for approximating IR metrics. The recent work by Bruch et al. \cite{RevisitingApproxNDCG} also shows its effectiveness based on deep neural networks. In particular, given the predicted score vector $\mathbf{y}$ obtained with the scoring function in response to a query, the rank position of the $i$-th document is approximated as
\begin{equation}
\label{Eq:QinRankApp}
\pi^{-}(i)=1+\sum_{j:j\neq i}\mathbb{I}\{y_{ij}<0\}
\end{equation}
\noindent where $y_{ij}=y_{i}-y_{j}$. $\mathbb{I}\{h\}$ is the indicator,
which is one if the condition $h$ is true and zero otherwise.
Furthermore, a smooth approximation of Eq-\ref{Eq:QinRankApp} is achieved by
substituting the indicator with a \textit{sigmoid} as follows:
\begin{equation}
\mathbb{I}\{y_{ij}<0\}\approx \sigma(-y_{ij}, \alpha)=\frac{1}{1+\exp(\alpha\cdot y_{ij})}
\end{equation}
where $\alpha>0$ is a knob that controls to what extent the sigmoid
fits the indicator. 

It is noteworthy that the metric scores being used as optimization objectives within the aforementioned approaches are approximations of the real metric scores. For example, Qin et al. \cite{QinApproximateNDCG} prove that the accuracy of their rank approximation can be given as follows:
\begin{equation}
|\pi^{-}(\mathbf{y})-\pi^{*}(\mathbf{y})|<\frac{|\mathbf{y}|-1}{\exp(\delta_{\mathbf{y}}\alpha)+1}
\end{equation}
\noindent where $\delta_{\mathbf{y}}=\min_{j:j\neq i}|y_{ij}|$, and $\pi^{*}(\mathbf{y})$ denotes the ground-truth ranks. Due to the inaccuracy of rank approximation, the obtained metric scores deviate from the real values.

\textbf{Neural ranking models} refer to the recent ranking methods~\cite{DSSM, CDSSM, DRMM, HuCNNMatching, PangAAAMatching, MatchSRNN} building upon neural networks. For example, the ranking models, such as DSSM \cite{DSSM} and CDSSM \cite{CDSSM}, map both queries and documents into the same semantic space
based on deep neural networks. The relevance score between a query and a document is assumed to be proportional to the cosine similarity between their corresponding vectors in the semantic space. The follow-up studies \cite{DRMM, HuCNNMatching, PangAAAMatching, MatchSRNN, Seq2Slate} look into the inherent characteristics of information retrieval. The DRMM model by Guo et al. \cite{DRMM} take into account more factors, such as query term importance, exact matching signals, and diverse matching requirement. The methods like \cite{HuCNNMatching, PangAAAMatching, MatchSRNN} first look at the local interactions between two texts, then design different network
architectures to learn more complicated interaction patterns for relevance matching. We refer the reader to \cite{IRJNeuralIR, IPMNeuralIR} for an overview of neural ranking models. Recently, there are a number of studies \cite{LinBERTDocR, RerankingBERT, CEDR} that explore how to fine-tune the bidirectional encoder representations from transformers (BERT) model \cite{BERT} to advance the ranking performance, e.g., for the passage re-ranking task\footnote{http://www.msmarco.org/leaders.aspx} based on MS MARCO. In this work, we use the data collection where the features of each query-document pair are prepared beforehand, and leave the exploration of using raw text queries and documents as a future work.
\section{Preliminaries}
\label{sec:Preli}
In this section, we introduce the general ranking framework following the Cranfield paradigm. We note that this framework is the same as or generalizes the ones employed in prior studies \cite{NDCGConsistency}.
\subsection{Cranfield Learning-to-Rank}
\label{subsec:L2R}
Let $\mathcal{Q}$ and $\mathcal{D}$ be the query space and the document
space, respectively, we use $\Phi:\mathcal{Q}\times\mathcal{D}\rightarrow\mathcal{Z}\coloneqq\mathbb{R}^{d}$
to denote the mapping function for generating a feature vector for
a document under a specific query context, where $\mathcal{Z}$ represents
the $d$-dimensional feature space. We use $\mathcal{T}\coloneqq\mathbb{R}_{\geq0}$
to denote the space of the ground-truth labels each document
receives. Thus for each query, we have a list of document feature
vectors $\mathbf{x}=(x_{1},...,x_{m})\in\mathcal{X}\coloneqq\mathcal{Z}^{m}$
and a corresponding list $\mathbf{y}^{\ast}=(y_{1}^{\ast},...,y_{m}^{\ast})\in\mathcal{Y}\coloneqq\mathcal{T}^{m}$
of ground-truth labels. The subscript $i$ like $x_{i}$ or $y_{i}^{\ast}$
denotes the $i$-position in the list. In practice, we get independently
and identically distributed (i.i.d) samples $\mathcal{S}=\{(\mathbf{x}_{j},\mathbf{y}_{j}^{\ast})\}_{j=1}^{n}$
from an unknown joint distribution $P(\cdot,\cdot)$ over $\mathcal{X}\times\mathcal{Y}$.
We use $f:\mathbf{x}\rightarrow\mathbb{R}^{m}$ parameterized
by $\theta\in\Theta$ to denote the real-valued scoring function,
which assigns each document a score. The scores of the documents associated
with the same query, i.e., $\mathbf{y}=f(\mathbf{x})=(f(x_{1}),f(x_{2}),...,f(x_{m}))$, are used to sort the documents.

A ranking $\pi$ on $m$ items $\mathbf{z}=[z_{1},...,z_{m}]$ is defined as an injection $\pi:\mathbf{z}\mapsto\{1,2,...,m\}$. Specifically, $\pi(i)$ / $\pi(z_i)$ yields the \textit{rank} of the $i$-th item in the list, where we think of higher positions with smaller rank values as more favorable. $\pi^{-1}(r)$ yields the
index of the item at rank $r$, and we have $\pi^{-1}(\pi(i))=i$ / $\pi^{-1}(\pi(z_i))=i$. Henceforth, we will use $\pi(i)$ and $\pi(z_i)$ interchangeably. We define the indexing operator $\llbracket\rrbracket$ as follows:
the zero-offset indices within $\llbracket\rrbracket$ specify the
elements to access, for instance, $z_{i}=\mathbf{z}\llbracket i\rrbracket$
and $[z_{i},z_{j},z_{k}]=\mathbf{z}\llbracket i,j,k\rrbracket$. \textit{Sorting in descending order} on a list
$\mathbf{z}$ is defined as: $\check{s}(\mathbf{z})$ yields the sorted
items such that $\check{s}(\mathbf{z})\llbracket j \rrbracket$ is monotonically decreasing
with the increasing of index $j$. $\check{s}^{-1}(\mathbf{z})$
yields the indices of the sorted items within the original list, and
we have $\check{s}(\mathbf{z})=\mathbf{z}
\llbracket\check{s}^{-1}(\mathbf{z})\rrbracket$.
Analogously, for \textit{sorting in ascending order}, we have $\hat{s}(\mathbf{z})=\mathbf{z}\llbracket\hat{s}^{-1}(\mathbf{z})\rrbracket$.
Assuming $\mathbf{z}=[1,3,5,4]$, Table \ref{Table:PermSort} illustrates how $\pi$ and
$\hat{s}$ work, which are further used in Section \ref{subsec:DiffIRMetrics} for deriving differentiable IR metrics.
\begin{table}[!htbp]	
	\caption{Permutation and sorting.}
	\label{Table:PermSort}
	\centering{}%
	\begin{tabular}{|c|c|c|c|c|}
		\hline 
		$\mathbf{z}$ & 1 & 3 & 5 & 4\tabularnewline
		\hline 
		$\pi(\mathbf{z})$ & 4 & 3 & 1 & 2\tabularnewline
		\hline 
		$\hat{s}(\pi(\mathbf{z}))$ & 1 & 2 & 3 & 4\tabularnewline
		\hline 
		$\hat{s}^{-1}(\pi(\mathbf{z}))$ & 2 & 3 & 1 & 0\tabularnewline
		\hline 
		$\mathbf{z}\llbracket\hat{s}^{-1}(\pi(\mathbf{z}))\rrbracket$ & 5 & 4 & 3 & 1\tabularnewline
		\hline 
	\end{tabular}
\end{table}

We measure the loss of ranking documents for a query using $f$ with
the \textit{rank-sensitive} loss function $\mathcal{R}(f(\mathbf{x}),\mathbf{y}^{\ast})$. The goal is to learn the optimal scoring function over a hypothesis space $\mathcal{F}$ of ranking functions that can \emph{minimize the expected risk} as below:
\begin{equation}
\min_{f\in\mathcal{F}}\Re(f)=\min_{f\in\mathcal{F}}\int_{\mathcal{X}\times\mathcal{Y}}\mathcal{R}(f(\mathbf{x}),\mathbf{y}^{\ast})dP(\mathbf{x},\mathbf{y}^{\ast})
\end{equation}

Typically, $\Re(f)$ is intractable to optimize directly and the joint distribution is unknown, we appeal to the \emph{empirical risk minimization} to approximate the expected risk, which is defined as follows:
\begin{equation}
\min_{f\in\mathcal{F}}\tilde{\Re}(f;\mathcal{S})=\min_{f\in\mathcal{F}}\frac{1}{n}\sum_{j=1}^{n}\mathcal{R}(f(\mathbf{x}_j),\mathbf{y}^{\ast}_j)
\end{equation}

Given the above general framework, one can design various ranking methods by deploying different loss functions to learn the parameters $\theta$ based on the training data. In the testing phase, the predicted ranking can be obtained efficiently by sorting the testing documents in descending order of their individual relevance scores.

\section{Differentiable IR Metrics}
In this section, we first provide a brief description on the widely used IR metrics. Then we propose the novel twin-sigmoid function, which enables to obtain the rank positions in a differentiable way. Finally, we show how to reformulate the IR metrics so as to bridge the gap between stochastic optimization and rank-sensitive metrics.
\subsection{Review on IR Metrics}
\label{subsec:IRMetric}
Given the predictions $\mathbf{y}=f(\mathbf{x})$ and the corresponding ground-truth
labels $\mathbf{y}^{*}$, we use $Y^{+}=\{i|y_{i}^{*}>0\}$ and $Y^{-}=\{j|y_{j}^{*}=0\}$ to represent the sets of relevant documents and non-relevant documents, respectively. We use $\mathbf{b}^{*}=\mathbb{I}\{\mathbf{y}^{*}>0\}$ with $b_{j}^{*}=\mathbb{I}\{y_{j}^{*}>0\}$
to represent the binarized ground-truth, and the cumulative sum on
$\mathbf{b}^{*}$ is given as $B_{k}^{*}=\sum_{j=1}^{k}b_{j}^{*}$. Based on the notation in Section \ref{subsec:L2R}, the scoring function $f$ induces
a ranking $\bar{\mathbf{y}}=\check{s}(\mathbf{y})$. 
The corresponding ground-truth labels are  $\mathbf{y}^{**}=\mathbf{y}^{*}\llbracket\check{s}^{-1}(\mathbf{y})\rrbracket$. Furthermore, we denote the binarized ground-truth labels as $\mathbf{b}^{**}=\mathbb{I}\{\mathbf{y}^{**}>0\}$
with $b_{j}^{**}=\mathbb{I}\{y_{j}^{**}>0\}$, and the cumulative
sum on $\mathbf{b}^{**}$ is given as $B_{k}^{**}=\sum_{j=1}^{k}b_{j}^{**}$. To evaluate the effectiveness of a scoring function, a number of IR
metrics have been proposed to emphasize the items that are ranked
at higher positions. In general, the IR metrics are computed
based on the list of ground-truth labels $\bar{\mathbf{y}}$ induced by $f$. For example, the binary-relevance IR metrics measure the performance of
a specific ranking model based on $\mathbf{b}^{**}$, such as precision and AP. The graded-relevance IR metrics measure the performance of a specific ranking model based on $\mathbf{y}^{**}$,
such as nDCG and ERR.

Precision@k measures the proportion
of relevant documents retrieved at a given truncation position, which is defined as:
\begin{equation}
Pre@k=\frac{1}{k}\sum_{j=1}^{k}b_{j}^{**}
\end{equation}
\noindent Here, $k$ denotes the truncation position.

Different from Precision@k that does not take into account the position
at which a document is ranked, \textbf{Average Precision} (AP) is a \textit{rank-sensitive}
metric, which builds upon Precision as follows:
\begin{equation}
AP=\frac{1}{|Y^{+}|}\sum_{j}b_{j}^{**}\times Pre@j
\end{equation}
\noindent Then Mean Average Precision (MAP) is defined
as the mean of AP scores over a set of queries.

\textbf{Normalized Discounted Cumulative Gain} (nDCG) \cite{FstnDCG} is a \textit{graded-relevance
	rank-sensitive} metric. The discounted cumulative
gain (DCG) of a ranked list is given as $DCG@k=\sum_{j=1}^{k}\frac{2^{y_{j}^{**}}-1}{\log_{2}(j+1)}$,
where $G_{j}=2^{y_{j}^{**}}-1$ is usually referred to as the gain
value of the $j$-th document.
We denote the maximum DCG value attained by the ideal ranking as $DCG^{*}$,
then normalizing DCG with $DCG^{*}$ gives nDCG as follows:
\begin{equation}
nDCG@k=\frac{DCG@k}{DCG^{*}@k}
\end{equation}

\textbf{Expected Reciprocal Rank} (ERR) \cite{ERR} is another popular graded-relevance rank-sensitive metric. Let $Pr(j)$ be the relevance probability of the document at rank
$j$. In accordance with the gain function for nDCG, the relevance
probability is commonly calculated as $Pr(j)=\frac{2^{y_{j}^{**}}-1}{2^{\max(\mathbf{y}^{**})}}$.
ERR interprets the relevance probability as the probability that the
user is satisfied with the document at a rank position. Thus the probability
that the user is dissatisfied with the documents at ranks from $1$
to $k$ is given as $Disp(1,k)=\prod_{i=1}^{k}(1-Pr(i))$. ERR is
then defined as
\begin{equation}
ERR@k=\sum_{j=1}^{k}\frac{Disp(1,j-1)\cdot Pr(j)}{j}
\end{equation}
\noindent Let $ERR^{*}$ the maximum ERR value attained by the ideal ranking, we have the normalized ERR as follows:
\begin{equation}
nERR@k=\frac{ERR@k}{ERR^{*}@k}
\end{equation}

\subsection{Rank Derivation}
\label{subsec:DSorting}
To address the aforementioned dilemma in Section-\ref{sec:Intro}, we propose the novel \textit{twin-sigmoid} function (referred to as $\sigma^{+}$). As its name indicates, the twin sigmoid consists of two ordinary sigmoid functions. Specifically, the \textit{forward sigmoid} function $\sigma_{f}^{+}(z,\alpha_{f})=\frac{1}{1+\exp(-\alpha_{f}\cdot z)}$ with a sufficient large scalar $\alpha_{f}$ is responsible for generating the output, namely $\sigma^{+}(z,\alpha_{f},\alpha_{b})=\sigma_{f}^{+}(z,\alpha_{f})$. The \textit{backward sigmoid} function $\sigma_{b}^{+}(z,\alpha_{b})=\frac{1}{1+\exp(-\alpha_{b}\cdot z)}$ with a small scalar $\alpha_{b}$ is responsible for generating the gradient for back-propagation, namely
\begin{equation}
\label{Eq:TwinsigmoidD1}
\frac{\partial\sigma^{+}(z,\alpha_{f},\alpha_{b})}{\partial z}=\alpha_{b}\cdot\sigma_{b}^{+}(z,\alpha_{b})\cdot[1-\sigma_{b}^{+}(z,\alpha_{b})]
\end{equation}

For instance, the blue and green curves in Figure \ref{subfig:twinSigmoid} show the output and derivative of the proposed twin-sigmoid, respectively. A closer look at Figure \ref{subfig:twinSigmoid} reveals that the twin-sigmoid has the following appealing properties: (1) The forward sigmoid with a sufficiently large $\alpha_{f}$ enables us to successfully mimic the indicator outputting either zero or one. (2) The backward sigmoid provides us the flexibility of back-propagating small gradients sidestepping the issue of gradient explosion. To further relieve the computational burden of twin-sigmoid, we omit the computation of the forward sigmoid with $\alpha_{f}\rightarrow\infty$, and directly output one for a positive input, zero for a negative input, and $0.5$ if the input is zero.

Armed with twin-sigmoid, the rank position
of the $i$-th document is given as
\begin{equation}
\label{Eq:VirSig4RankDerivation}
\pi^{+}(i)=1+\sum_{j:j\neq i} 1- \sigma^{+}(y_{ij}, \alpha_{f},\alpha_{b})
\end{equation}

To overcome the impact of score ties on rank approximation with Eq-\ref{Eq:VirSig4RankDerivation}, we appeal to the random shuffle flavored strategy. 
Given the prediction $\mathbf{y}$, we denote its pairwise comparison matrix as $A=[y_{ij}]$.
Let $\mathbf{p}$ be a random permutation of integers from $1$ to $m$, and $p_i$ is the $i$-th integer. We denote its pairwise comparison matrix as $P=[p_{ij}]$ with $p_{ij}=p_i-p_j$. Finally, the binary matrix $\ddot{P}$ with $\ddot{P}_{ij}=\mathbb{I}\{p_{ij}>0\}$ is used. Specifically, for a non-diagonal zero element $y_{ij}$ of $A$, the output of $\sigma^{+}(y_{ij})$ (which is $0.5$) is further rectified as $\ddot{P}_{ij}$\footnote{In this work, the rectified value is obtained as: $\sigma^{+}(y_{ij}) + 0.5$ if $\ddot{P}_{ij}$ is one, and $\sigma^{+}(y_{ij}) - 0.5$ otherwise.}. In a nutshell, relying on the binarized pairwise comparison matrix of a random permutation of distinct integers, we break tie predictions in $\mathbf{y}$ randomly.

To illustrate the effectiveness of rank prediction, we compare our proposed method $\pi^{+}$ against the method $\pi^{-}$ in \cite{QinApproximateNDCG} based on two synthetic datasets, which are drawn from a uniform distribution on the interval $[0, 1)$. In particular, each dataset consists of $v_1$ vectors (denoted as $\{\mathbf{z}\}$) and each vector consists of $v_2$ values. Here $v_1$ and $v_2$ are used to mimic the number of queries and the number of documents, respectively. We use the average $L1$ loss between the predicted ranks and the ground-truth ranks to measure the accuracy, which is defined as $L1(\pi;\{\mathbf{z}\})=\frac{1}{v_{1}}\sum_{k}|\pi(\mathbf{z}_{k})-\pi^{*}(\mathbf{z}_{k})|$.

\begin{table}[!htbp]	
	\caption{The average L1 loss of $\pi^{-}$ for rank approximation.}
	\label{Table:RankAppCmp}
	\centering{}%
	\resizebox{0.48\textwidth}{!}{
		\begin{tabular}{|c|c|c|c|c|c|c|}
			\hline 
			$\sigma$ & 1 & 10 & 100 & 1,000 & 10,000 & 100,000\tabularnewline
			\hline 
			$(v_1=100, v_2=123)$ & 2,866.94 & 350.25 & 68.36 & 16.45 & 2.20 & 0.23\tabularnewline
			\hline 
			$(v_1=100, v_2=1000)$ & 189,401.48 & 17,600.48 & 1,671.72 & 488.01 & 112.80 & 13.67\tabularnewline
			\hline 
		\end{tabular}
	}
\end{table}
\begin{table}
	\caption{The average L1 loss of $\pi^{+}$ for rank approximation.}
	\label{Table:RankAppCmp2}
	\begin{centering}
		\resizebox{0.48\textwidth}{!}{
			\begin{tabular}{|c|c|c|}
				\hline 
				$\pi^{+}$ & ($v_{1}=100$, $v_{2}=123$) & ($v_{1}=100$, $v_{2}=1000$)\tabularnewline
				\hline 
				Eq-\ref{Eq:VirSig4RankDerivation} & 0 & 0.03\tabularnewline
				\hline 
				Eq-\ref{Eq:VirSig4RankDerivation} plus tie breaking & 0 & 0\tabularnewline
				\hline 
			\end{tabular}
		}
		\par\end{centering}
\end{table}

From Table \ref{Table:RankAppCmp}, we can observe that: with a sufficiently large $\sigma$, the L1 loss of $\pi^{-}$ for rank approximation is significantly decreased. However, the derivative of the composing sigmoid in $\pi^{-}$ would become quite large, which makes it hard to perform stochastic optimization using neural networks. From Table \ref{Table:RankAppCmp2}, we can see that: armed with twin-sigmoid, $\pi^{+}$ achieves significantly lower L1 loss, which results from the impact of score ties. By resorting to the strategy of tie breaking, $\pi^{+}$ is able to obtain the correct ranks rather than approximation. This is also why we tried different activation functions in the output layer in Section \ref{sec:ResAndAna}, where the differences between $\pi^{+}$ and $\pi^{-}$ for direct metric optimization are also demonstrated.
\subsection{Differentiable Formulation of IR Metrics}
\label{subsec:DiffIRMetrics}
Given the predictions $\mathbf{y}=f(\mathbf{x})$, the ranks of each document are given as $\mathbf{r}=\pi^{+}(\mathbf{y})$. Furthermore, we sort the predicted ranks in ascending order, namely
$\bar{\mathbf{r}}=\hat{s}(\pi^{+}(\mathbf{y}))$, the values of
$\bar{\mathbf{r}}$ essentially are $[1,2,...,m]$. The corresponding ground-truth labels will be $\mathbf{y}^{**}=\mathbf{y}^{*}[\hat{s}^{-1}(\pi^{+}(\mathbf{y}))]$.
In fact, $\mathbf{y}^{*}[\hat{s}^{-1}(\pi^{+}(\mathbf{y}))]$ and
$\mathbf{y}^{*}[\check{s}^{-1}(\mathbf{y})]$ induce the same result. It is noteworthy that the sorting operation has no impact on the differentiability of both $\mathbf{r}$
and $\bar{\mathbf{r}}$. When reformulating the IR metrics, $\bar{\mathbf{r}}$
can be directly used as sequential differentiable ranks from $1$
to $m$. Plugging the differentiable ranks $\mathbf{r}$ or $\bar{\mathbf{r}}$ into the IR metrics in Section \ref{subsec:IRMetric} yields their corresponding differentiable formulations.

Specifically, differentiable precision is formulated as follows:
\begin{equation}
\label{Eq:VirPre}
\widehat{Pre}@k=\frac{1}{k}\sum_{i=1}^{k}b_{i}^{**}\frac{i}{\bar{r_{i}}}
\end{equation}
\noindent Note that the value of $\bar{r_{i}}$ is $i$, but it is differentiable. Then we get the differentiable AP as follows:
\begin{equation}
\label{Eq:VirAP}
\widehat{AP}=\frac{1}{|Y^{+}|}\sum_{k=1}^{m}b_{k}^{**}\frac{1}{k}\sum_{i=1}^{k}b_{i}^{**}\frac{i}{\bar{r_{i}}}
\end{equation}

By replacing the numerical rank position with the differentiable rank prediction, the differentiable nDCG can be given as:
\begin{equation}
\label{Eq:VirNDCG}
\widehat{nDCG}=\frac{1}{DCG^{*}}\sum_{k}\frac{2^{y_{k}^{*}}-1}{\log_{2}(r_{k}+1)}
\end{equation}

Analogously, the differentiable nERR can be given as:
\begin{equation}
\label{Eq:VirNERR}
\widehat{nERR}@k=\frac{1}{ERR^{*}@k}\sum_{j=1}^{k}\frac{\prod_{i=1}^{j-1}(1-\frac{2^{y_{i}^{**}}-1}{2^{\max(\mathbf{y}^{**})}})\cdot\frac{2^{y_{j}^{**}}-1}{2^{\max(\mathbf{y}^{**})}}}{\bar{r_{j}}}
\end{equation}
\section{Stochastic Optimization of IR Metrics}
\label{sec:OptIRMetrics}
In this section, by carrying out an in-depth analysis of the gradients, we first discover two potential limitations inherent with the vanilla direct optimization of IR metrics. To break the limitations, then we propose different methods by incorporating ground-truth labels and virtual gradient.

With the aforementioned reformulation, the widely used IR metrics become differentiable with respect to the parameters $\theta$ of the scoring function. By defining the \textit{negative metric score} as the ranking loss, one can appeal to many optimization algorithms, such as mini-batch gradient descent for loss minimization. We denote this type of direct metric optimization as \textit{type1}. Using nDCG as an example, a closer look at Eq-\ref{Eq:VirNDCG} reveals that the nDCG score (other metrics are the same) is merely derived from relevant documents since the gain value of a non-relevant document is zero. For a specific query, the predicted rank position of the document $x_{i}$ is $r_{i}=1+\sum_{j:j\neq i}1-\sigma^{+}(y_{ij},\alpha_{f},\alpha_{b})$, the gradients of $r_{i}$ with respect to the predictions of documents $x_{i}$ and $x_{j}$ the can be computed as
\begin{equation}
\frac{\partial r_{i}}{\partial y_{i}}=\sum_{j:j\neq i}-\alpha_{b}\cdot \sigma^{+}_{b}(y_{ij}, \alpha_{b})\cdot[1-\sigma^{+}_{b}(y_{ij}, \alpha_{b})]
\end{equation}
\begin{equation}
\frac{\partial r_{i}}{\partial y_{j}}=\alpha_{b}\cdot \sigma^{+}_{b}(y_{ij}, \alpha_{b})\cdot[1-\sigma^{+}_{b}(y_{ij}, \alpha_{b})]
\end{equation}

Assuming that the document $x_{i}$ is relevant, the gradients of the ranking loss (i.e., minus nDCG) w.r.t. the predictions $y_i$ and $y_j$ through $r_{i}$ are

\begin{equation}
\label{Eq:GradYi}
	\begin{split}
	&\frac{\partial\mathcal{R}_{nDCG}(f(\mathbf{x}),\mathbf{y}^{*})}{\partial y_{i}}=
	\sum_{k:y_{k}^{*}>0}\frac{2^{y_{k}^{*}}-1}{DCG^{*}}\cdot\frac{-1}{(\log_{2}(r_{k}+1))^{2}}\cdot\frac{1}{r_{k}+1}\cdot\frac{\partial r_{k}}{\partial y_{i}}\\
	&=\frac{2^{y_{i}^{*}}-1}{DCG^{*}}\cdot\frac{-1}{(\log_{2}(r_{i}+1))^{2}}\cdot\frac{1}{r_{i}+1}\cdot\frac{\partial r_{i}}{\partial y_{i}}
	+\sum_{k:k\neq i;y_{k}^{*}>0}\frac{2^{y_{k}^{*}}-1}{DCG^{*}}\cdot\frac{-1}{(\log_{2}(r_{k}+1))^{2}}\cdot\frac{1}{r_{k}+1}\cdot\frac{\partial r_{k}}{\partial y_{i}}
	\end{split}
\end{equation}


The above gradient is further back-propagated to update the parameters of the scoring function in a way like: $\theta_{t}=\theta_{t-1}+\eta\cdot\nabla_{\theta}\mathcal{R}_{nDCG}(f(\mathbf{x}),\mathbf{y}^{*})$, where $\eta$ is a positive learning rate.

Unfortunately, there are some potential problems inherent with direct optimization of IR metrics following Eq-\ref{Eq:GradYi}. First, a closer look at Eq-\ref{Eq:GradYi} reveals that: the relevant document $x_{i}$ always get \textit{forces} from all other relevant documents $\{d_{k}\}$ so as to tune the scoring function for $y_{i}-y_{k}<0$. As a result, for highly relevant documents, this is not desirable since we hope in each iteration the optimization direction is consistent with the ground-truth labels. We name it as \textit{the problem of optimization inconsistency}. To overcome this problem, we
propose to modify the back-propagating gradients of $\sigma^{+}$ by incorporating the ground-truth labels as follows:
\begin{equation}
\label{Eq:type2}
\frac{\partial\sigma^{+}(y_{ij},\alpha_{f},\alpha_{b})}{\partial y_{ij}}=u_{ij}\cdot\alpha_{b}\cdot\sigma^{+}_{b}(y_{ij},\alpha_{b})\cdot[1-\sigma^{+}_{b}(y_{ij},\alpha_{b})]
\end{equation}
\noindent where $u_{ij}$ is a signal determined via ground-truth labels. Its value is $1$ if $y_{i}^{*}>y_{j}^{*}$, $0$ if $y_{i}^{*}=y_{j}^{*}$, and $-1$
if $y_{i}^{*}<y_{j}^{*}$. We denote the type of direct metric optimization based on Eq-\ref{Eq:type2} as \textit{type2}.

Second, from Figure \ref{subfig:derivativeOfVanillaSigmoid}, we can observe that the gradient of a sigmoid function becomes larger when the absolute value of the input approaches zero. The gradient decreases quickly when the absolute value of the input becomes larger. Supposing that $x_{i}$ is a relevant document, and $x_{j}$ is a non-relevant document, the predictions $y_i$ and $y_j$ at a certain optimization iteration have $y_{i}-y_{j}\ll0$. According to Figure \ref{subfig:twinSigmoid}, $y_{i}-y_{j}\ll0$ leads to a small gradient that is close to $0$. As a result, correcting the predictions $y_i$ and $y_j$ for better performance becomes difficult within the current optimization iteration. This is again not
desirable since we hope to quickly correct the wrongly ranked pairs in each iteration. We name it as \textit{the problem of optimization difficulty}. To overcome this problem, we further modify the back-propagating gradients of $\sigma^{+}$ as follows:

\begin{equation}
\label{Eq:type3}
\frac{\partial\sigma^{+}(y_{ij},\alpha_{f},\alpha_{b})}{\partial y_{ij}}=\begin{cases}
2\alpha_b\cdot[1-\sigma^{+}_{b}(y_{ij},\alpha_b)] & u_{ij}=1\\
0 & u_{ij}=0\\
-2\alpha_b\cdot\sigma^{+}_{b}(y_{ij},\alpha_b) & u_{ij}=-1
\end{cases}
\end{equation}

The motivation behind this modification is to overcome the aforementioned training difficulty by amplifying the training signal. The red curve in Figure \ref{subfig:twinSigmoid}  plots the gradient variation with respect to $\alpha_b=1.0$ and $u_{ij}=1.0$. We denote the type of direct metric optimization based on Eq-\ref{Eq:type3} as \textit{type3}.

The proposed methods for metric optimization are approximately smooth but not convex\footnote{Given a large number of queries, the average IR metric becomes approximately smooth, which makes it possible to compute an empirical gradient \cite{EmpiricalSmoothMetric}.}, there may be many local optima during training. Thus we used 5-fold cross validation strategy to report the average performance. We plan to explore other global optimization methods in the future. Regarding the efficiency, at training time, the proposed framework for direct metric optimization has the same time complexity as ApproxNDCG \cite{QinApproximateNDCG} and WassRank \cite{TaoWSDM2019}, which is of order $\mathcal{O}(m^{2})$ for a single query. At test time, there is no difference from any other method, namely sorting the documents in descending order of predicted scores.
\section{Experimental Setup}
In this section, we describe the experimental setup. We first introduce the data collection and the way of evaluation. We then describe the configuration of each method to be evaluated.
\subsection{Dataset}
We used the benchmark dataset, MSLR-WEB30K, which is the largest one among the LETOR datasets. Each query-document pair is represented with a feature vector. The ground truth is a multiple-level relevance judgment, which takes $5$ values from $0$ (irrelevant) to $4$ (perfectly relevant). The basic statistics are: the number of queries is $30,295$, the number of documents is $3,749,144$, the number of features is $136$ and the average number of relevant documents per query is $60$. For more detailed information, e.g., the feature description, we refer readers to the overview paper \cite{LETORIR}. We use nDCG and MAP to measure the performance. We report the results with different cutoff values $1$, $3$, $5$, $10$ and $20$ to show the performance of each method at different positions. As discussed in Section \ref{subsec:IRMetric}, different from MAP that merely considers the rank position, nDCG takes into account both the rank position and the relevance level. Thus \textit{nDCG is used as the main effectiveness measure} in this work. We observe that the results in terms of nERR are consistent with nDCG, which are not included due to space constraints.

We note that the previous studies \cite{RevisitingApproxNDCG, StochasticTreatmentRF, LambdaLossFramework} just used a single fold (i.e., Fold1) for the experimental evaluation. To reduce the possible impact of overfitting on performance comparison, we use all the five folds and perform 5-fold cross validation in this work. In particular, the dataset is randomly partitioned into five equal sized subsets. In each fold, three subsets are used as the training data, the remaining two subsets are used as the validation data and the testing data, respectively. We use the training data to learn the ranking model, use the validation data to select the hyper parameters based on nDCG@5, and use the testing data for evaluation. Finally, we report the ranking performance based on the averaged evaluation scores across five folds with $100$ epochs. 
\subsection{Baselines and Model Configuration} In our experiments, a number of representative approaches are used as our baselines: (1) ListNet \cite{ListNet}, ListMLE \cite{ListMLE} and WassRank \cite{TaoWSDM2019} are adopted to represent the approaches that ignore evaluation metrics during the training process. (2) LambdaMART \cite{LambdaMART} is empirically shown to be the \textit{state-of-the-art approach} based on the technique of gradient boosting decision tree (GBDT). In this work we use the implementation included in LightGBM \cite{LightGBM}, which is referred to as LambdaMART(L). (3) ApproxNDCG \cite{QinApproximateNDCG} is adopted to represent the main baseline approach that directly optimizes an evaluation metric for ranking. The recent work by Bruch et al. \cite{RevisitingApproxNDCG} showed its effectiveness based on neural networks again. The other approaches  \cite{YueSVMAP, SoftRank, AdaRank} are not included since they underperform ApproxNDCG according to the work by Qin et al. \cite{QinApproximateNDCG}. 

We implemented and trained all the proposed methods and baseline approaches (except LambdaMART) using PyTorch v1.3, where one Nvidia Titan RTX GPU with 24 GB memory is used\footnote{We will release the source code to enable reproduction and extension of our work.}. We used the L2 regularization with a decaying rate of $1\times10^{-3}$ and the Adam optimizer with a learning rate of $1\times10^{-3}$. We used a simple $5$-layer feed-forward neural network, where the size of a hidden layer is set as $100$. We adopted two types of activation functions \textit{ReLU} and \textit{CELU}. We also explored whether to apply an activation function in the last layer or not. In total, we tried four different architecture settings, which are referred to as R5, CE5, R4.L and CE4.L, respectively. As an example, R4.L refers to that ReLU is used in the first four layers, and the last layer is linear without using any activation function. Given the raw features per query-document pair, they are normalized using the \textit{z-score} method at a query level. We further use batch normalization between consecutive layers. For all the proposed methods and the baseline approaches based on neural networks, we apply the same framework (e.g., the scoring function and the tuning strategy) except the component of loss function. This enables us to conduct a fair comparison when investigating the impact of a specific component on the performance.

For ListNet, the ranking loss is computed based on the top-1 approximation as in the original paper \cite{ListNet}, namely each element of the probability vector represents the probability of the corresponding document being ranked at the top-1 position. For WassRank, the suggested parameter configuration by \cite{TaoWSDM2019} is used. Following the recent studies \cite{RevisitingApproxNDCG, StochasticTreatmentRF}, for ApproxNDCG, the parameter $\alpha$ is set as $10$. According to \cite{StochasticTreatmentRF}, for LambdaMART(L), the parameters are set as: learning rate is $0.05$, num\_leaves is $400$, min\_data\_in\_leaf is $50$, and min\_sum\_hessian\_in\_leaf is set to $200$. We use nDCG@5 to select the best models on validation sets by fixing early stopping round to $200$ up to $1000$ trees. For the proposed framework, we take Pre, AP, nDCG and nERR@10 as the optimization objectives, respectively. According to the type$k$ ($k=1,2,3$) of metric optimization in Section \ref{sec:OptIRMetrics}, the methods are referred to as Pre-type$k$, AP-type$k$, nDCG-type$k$ and nERR@10-type$k$, respectively. During the optimization, $\alpha_b$ are set as $1.0$, and the tie breaking strategy is deployed.

\section{Results and Analysis}
\label{sec:ResAndAna}
In this section, we report the experimental results and conduct detailed analysis. Particularly, we want to show how effective are the proposed methods by directly optimizing a specific evaluation metric and shed some light on why it is able to achieve improved performance. In the following, we first compare the overall performance, and then examine the training process of each method.
\subsection{Overall Performance}
\label{subsec:OverallCmp}
In Table \ref{Table:BaselineResults}, Table \ref{Table:Results_VirSig}, Table \ref{Table:Results_SignVirSig} and Table \ref{Table:Results_SignVirInCE}, we show the overall performance of the baseline approaches, and the proposed methods, respectively. Within each table, the best result of each setting is indicated in bold, where the superscript ${\ast}$ indicates statistically significant difference when compared to the best result based on the Wilcoxon signed-rank test with $p<0.01$.

We first look at the performance of baseline approaches in Table \ref{Table:BaselineResults}. First, we can observe that LambdaMART(L) achieves significantly better performance than other baseline approaches in terms of nDCG. The reasons are that: The objective optimized by LambdaMART is a coarse upper bound of nDCG \cite{LambdaLossFramework}. Benefiting from GBDT in the form of an ensemble of weak prediction models and the algorithmic and engineering optimizations of LightGBM, LambdaMART(L) shows more promising results. Compared with the other baseline approaches, ListMLE performs the worst in terms of nDCG. On the contrary, ListMLE performs the best in terms of MAP. The main reasons are that: (1) Though ListMLE has been proved to be consistent with permutation level 0-1 loss \cite{ListMLE}, \textit{it does not mean consistency with nDCG}. (2) ListMLE is defined in a top-down style that seems to reflect the position importance in ranking. According to \cite{PListMLE}, the decomposition of probability in ListMLE is not unique due to the chain rule of probability. Among the baseline approaches based on neural networks, ApproxNDCG performs the best in terms of nDCG, which shows that approximating the target evaluation metric as the loss function is a more appropriate choice than the surrogate loss functions used by ListMLE, ListNet and WassRank. Our observation is also consistent with the results reported by \cite{RevisitingApproxNDCG}.

We next look at the performance of the proposed methods of type1 for direct metric optimization in Table \ref{Table:Results_VirSig}. We can observe that Pre-type1, AP-type1 and nERR10-type1 show poor performance when compared with the baseline approaches. We believe that this is primarily because of the aforementioned problems of optimization inconsistency and optimization difficulty in Section \ref{sec:OptIRMetrics}. However, nDCG-type1 shows comparative performance to ApproxNDCG and outperforms ListMLE, ListNet and WassRank. Note that the main difference between nDCG-type1 and ApproxNDCG is the tuning knob $\alpha$. To overcome the limitations underlying metric optimization of type1, we propose to explicitly modify the gradient computation, i.e., Eq-\ref{Eq:type2} and Eq-\ref{Eq:type3}. Table \ref{Table:Results_SignVirSig} and Table \ref{Table:Results_SignVirInCE} describe the effectiveness of metric optimization based on the type2 and type3 strategies, respectively. By comparing Table \ref{Table:Results_VirSig} and Table \ref{Table:Results_SignVirSig}, we can find that Pre-type2, AP-type2 and nERR10-type2 show improved performance, especially AP-type2. Therefore, incorporating the indicator of ground-truth comparison result helps to improve metric optimization. Furthermore, from Table \ref{Table:Results_SignVirInCE}, we can observe that Pre-type3, AP-type3, nERR10-type3 and nDCG-type3 show significantly improved performance, especially compared with the corresponding performance in Table \ref{Table:Results_VirSig}. 

A joint look at Table \ref{Table:BaselineResults}, Table \ref{Table:Results_VirSig}, Table \ref{Table:Results_SignVirSig} and Table \ref{Table:Results_SignVirInCE} reveals that: (1) Compared with the baseline approaches that ignore the evaluation metric during the optimization process, direct IR metric optimization based on the vanilla sigmoid is able to obtain relatively better performance, such as ApproxNDCG and nDCG-type1. (2) However, there are inherent limitations underlying the vanilla sigmoid due to its symmetric bell-curve gradient computation, namely optimization inconsistency and optimization difficulty. Thanks to the proposed gradient modifications of type2 and type3, we can overcome these limitations. The improved results in Table \ref{Table:Results_SignVirSig} and Table \ref{Table:Results_SignVirInCE} echo the above analysis. (3) Intuitively, compared with the optimization of precision and AP, direct optimization of nDCG should lead to better performance in terms of nDCG. The results in Table \ref{Table:Results_VirSig} look consistent, while the results in Table \ref{Table:Results_SignVirSig} and Table \ref{Table:Results_SignVirInCE} seem counterintuitive. A reasonable explanation is that the effects of gradient modification (Eq-\ref{Eq:type2} and Eq-\ref{Eq:type3}) on precision and AP are more pronounced. (4) Regarding the activation function in the last layer, ReLU generally leads lower performance. This is
attributable to score ties when conducting pairwise comparisons for deriving the rank positions, because ReLU performs a threshold operation that any input value less than zero is set to zero. Therefore, for direct metric optimization based on either sigmoid or twin-sigmoid, careful examinations of the activation function in the last layer are highly recommended.

\begin{table}
	\caption{The performance of baseline approaches on MSLRWEB30K.}
	\label{Table:BaselineResults} 
	\begin{centering}
		\resizebox{\textwidth}{!}{
			\begin{tabular}{|l|c|c|c|c|c|c|c|c|c|c|}
				\hline 
				& nDCG@1 & nDCG@3 & nDCG@5 & nDCG@10 & nDCG@20 & MAP@1 & MAP@3 & MAP@5 & MAP@10 & MAP@20\tabularnewline
				\hline
				\multirow{1}{*}{ListMLE (CE5)} & $0.4637^{\ast}$ & $0.4474^{\ast}$ & $0.4507^{\ast}$ & $0.4668^{\ast}$ & $0.4881^{\ast}$ & \textbf{0.2888} & \textbf{0.2838} & \textbf{0.2839} & \textbf{0.2883} & \textbf{0.2947}\tabularnewline
				\hline 
				\multirow{1}{*}{ListNet (R4.L)} & $0.4665^{\ast}$ & $0.4495^{\ast}$ & $0.4534^{\ast}$ & $0.4708^{\ast}$ & $0.4942^{\ast}$ & $0.2710^{\ast}$ & $0.2620^{\ast}$ & $0.2620^{\ast}$ & $0.2666^{\ast}$ & $0.2751^{\ast}$\tabularnewline
				\hline 
				\multirow{1}{*}{ApproxNDCG (R4.L)} & $0.4768^{\ast}$ & $0.4538^{\ast}$ & $0.4554^{\ast}$ & $0.4691^{\ast}$ & $0.4890^{\ast}$ & $0.2829^{\ast}$ & $0.2757^{\ast}$ & $0.2754^{\ast}$ & $0.2779^{\ast}$ & $0.2832^{\ast}$\tabularnewline
				\hline 
				\multirow{1}{*}{WassRank (CE4.L)} & $0.4684^{\ast}$ & $0.4460^{\ast}$ & $0.4473^{\ast}$ & $0.4611^{\ast}$ & $0.4814^{\ast}$ & $0.2814^{\ast}$ & $0.2748^{\ast}$ & $0.2737^{\ast}$ & $0.2765^{\ast}$ & $0.2824^{\ast}$\tabularnewline
				\hline
				LambdaMART(L) & \textbf{0.4933} & \textbf{0.4743} & \textbf{0.4776} & \textbf{0.4948} & \textbf{0.5166} & 0.2874 & $0.2829^{\ast}$ & $0.2835^{\ast}$ & $0.2879^{\ast}$ & $0.2941^{\ast}$\tabularnewline
				\hline 
				
			\end{tabular}
		}
		\par\end{centering}
\end{table}

\begin{table}
	\caption{Direct optimization of precision, average precision, nERR@10 and nDCG based on the strategy \textit{type1}.}
	\label{Table:Results_VirSig}
	\begin{centering}
		\resizebox{\textwidth}{!}{
			\begin{tabular}{|l|c|c|c|c|c|c|c|c|c|c|}
				\hline 
				& nDCG@1 & nDCG@3 & nDCG@5 & nDCG@10 & nDCG@20 & MAP@1 & MAP@3 & MAP@5 & MAP@10 & MAP@20\tabularnewline
				\hline 
				\multirow{1}{*}{Pre-type1 (R4.L)} & $0.4440^{\ast}$ & $0.4302^{\ast}$ & $0.4363^{\ast}$ & $0.4535^{\ast}$ & $0.4753^{\ast}$ & 0.2863 & 0.2812 & \textbf{0.2821} & \textbf{0.2865} & \textbf{0.2932}\tabularnewline
				\hline 
				\multirow{1}{*}{AP-type1 (R4.L)} & $0.4424^{\ast}$ & $0.4288^{\ast}$ & $0.4337^{\ast}$ & $0.4499^{\ast}$ & $0.4713^{\ast}$ & \textbf{0.2866} & \textbf{0.2813} & 0.2816 & 0.2850 & 0.2909\tabularnewline
				\hline 
				\multirow{1}{*}{nERR@10-type1 (R5)} & $0.4609^{\ast}$ & $0.4412^{\ast}$ & $0.4418^{\ast}$ & $0.4518^{\ast}$ & $0.4632^{\ast}$ & $0.2795^{\ast}$ & $0.2716^{\ast}$ & $0.2699^{\ast}$ & $0.2684^{\ast}$ & $0.2653^{\ast}$\tabularnewline
				\hline 
				\multirow{1}{*}{nDCG-type1 (R4.L)} & \textbf{0.4754} & \textbf{0.4556} & \textbf{0.4565} & \textbf{0.4705} & \textbf{0.4906} & $0.2811^{\ast}$ & $0.2757^{\ast}$ & $0.2751^{\ast}$ & $0.2775^{\ast}$ & $0.2834^{\ast}$\tabularnewline
				\hline 
			\end{tabular}
		}
		\par\end{centering}
\end{table}

\begin{table}
	\caption{Direct optimization of precision, average precision, nERR@10 and nDCG based on the strategy \textit{type2}.}
	\label{Table:Results_SignVirSig}
	\begin{centering}
		\resizebox{\textwidth}{!}{
			\begin{tabular}{|l|c|c|c|c|c|c|c|c|c|c|}
				\hline 
				& nDCG@1 & nDCG@3 & nDCG@5 & nDCG@10 & nDCG@20 & MAP@1 & MAP@3 & MAP@5 & MAP@10 & MAP@20\tabularnewline
				\hline 
				\multirow{1}{*}{Pre-type2 (R4.L)} & 0.4741 & 0.4551 & $0.4587^{\ast}$ & \textbf{0.4764} & \textbf{0.4986} & 0.2841 & 0.2790 & \textbf{0.2801} & \textbf{0.2851} & \textbf{0.2920}\tabularnewline
				\hline 
				\multirow{1}{*}{AP-type2 (R5)} & \textbf{0.4764} & \textbf{0.4573} & \textbf{0.4606} & 0.4757 & 0.4954 & \textbf{0.2842} & \textbf{0.2791} & \textbf{0.2801} & 0.2843 & 0.2900\tabularnewline
				\hline 
				\multirow{1}{*}{nERR@10-type2 (R5)} & $0.4694^{\ast}$ & $0.4455^{\ast}$ & $0.4451^{\ast}$ & $0.4557^{\ast}$ & $0.4685^{\ast}$ & $0.2795^{\ast}$ & $0.2717^{\ast}$ & $0.2702^{\ast}$ & $0.2705^{\ast}$ & $0.2698^{\ast}$\tabularnewline
				\hline 
				\multirow{1}{*}{nDCG-type2 (R4.L)} & 0.4753 & $0.4551^{\ast}$ & $0.4568^{\ast}$ & 0.4709 & 0.4916 & $0.2791^{\ast}$ & $0.2729^{\ast}$ & $0.2728^{\ast}$ & $0.2755^{\ast}$ & $0.2817^{\ast}$\tabularnewline
				\hline 
			\end{tabular}
		}
		\par\end{centering}
\end{table}

\begin{table}
	\caption{Direct optimization of precision, average precision, nERR@10 and nDCG based on the strategy \textit{type3}.}
	\label{Table:Results_SignVirInCE}
	\begin{centering}
		\resizebox{\textwidth}{!}{
			\begin{tabular}{|l|c|c|c|c|c|c|c|c|c|c|}
				\hline 
				& nDCG@1 & nDCG@3 & nDCG@5 & nDCG@10 & nDCG@20 & MAP@1 & MAP@3 & MAP@5 & MAP@10 & MAP@20\tabularnewline
				\hline 
				\multirow{1}{*}{Pre-type3 (R5)} & $0.4807^{\ast}$ & $0.4589^{\ast}$ & 0.4628 & 0.4800 & \textbf{0.5013} & \textbf{0.2852} & \textbf{0.2796} & \textbf{0.2807} & \textbf{0.2856} & \textbf{0.2919}\tabularnewline
				\hline 
				\multirow{1}{*}{AP-type3 (CE4.L)} & \textbf{0.4855} & \textbf{0.4630} & \textbf{0.4646} & \textbf{0.4805} & 0.5012 & 0.2849 & 0.2790 & $0.2789^{\ast}$ & $0.2832^{\ast}$ & $0.2887^{\ast}$\tabularnewline
				\hline 
				\multirow{1}{*}{nERR@10-type3 (CE5)} & 0.4834 & $0.4553^{\ast}$ & $0.4551^{\ast}$ & $0.4680^{\ast}$ & $0.4869^{\ast}$ & $0.2792^{\ast}$ & $0.2708^{\ast}$ & $0.2694^{\ast}$ & $0.2709^{\ast}$ & $0.2754^{\ast}$\tabularnewline
				\hline 
				\multirow{1}{*}{nDCG-type3 (CE4.L)} & 0.4833 & $0.4598^{\ast}$ & $0.4604^{\ast}$ & $0.4744^{\ast}$ & $0.4951^{\ast}$ & $0.2788^{\ast}$ & $0.2720^{\ast}$ & $0.2708^{\ast}$ & $0.2734^{\ast}$ & $0.2790^{\ast}$\tabularnewline
				\hline 
			\end{tabular}
		}
		\par\end{centering}
\end{table}
\subsection{Examination of Training Process}
\label{subsec:TrainProcess}

\begin{figure}[!htbp]
	\begin{centering}
		\subfigure[WassRank.\label{subfig:WassRank}]
		{
			\centering{}\includegraphics[width=.24\textwidth, totalheight=1.2in]{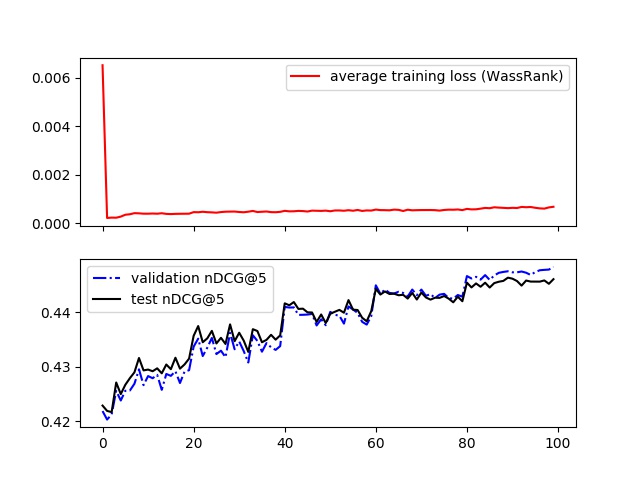}}
		\subfigure[ListNet.\label{subfig:ListNet}]
		{
			\centering{}\includegraphics[width=.24\textwidth, totalheight=1.2in]{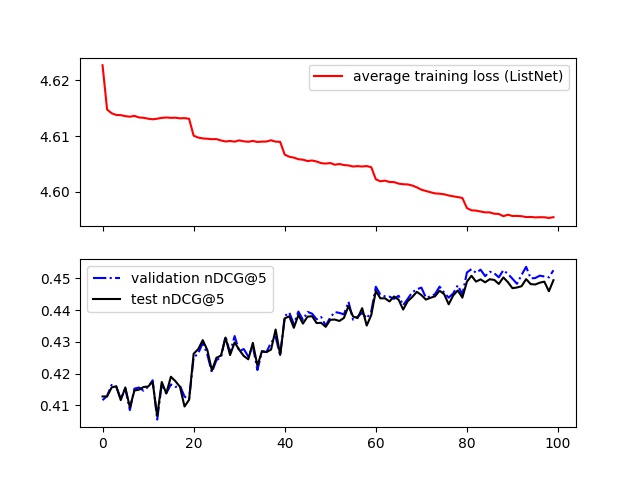}}		
		\subfigure[ListMLE.\label{subfig:ListMLE}]
		{
			\centering{}\includegraphics[width=.24\textwidth, totalheight=1.2in]{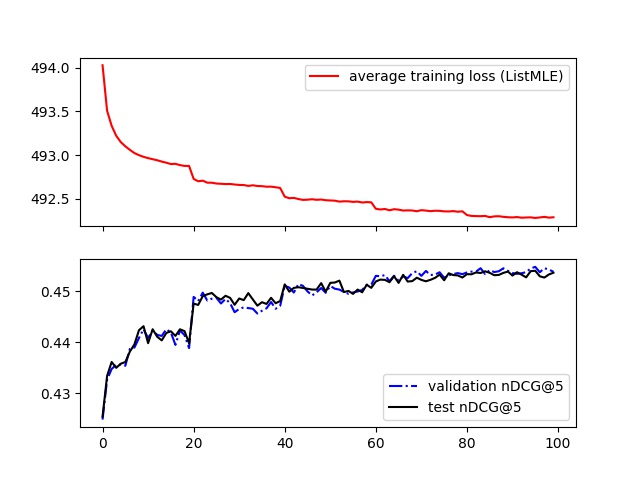}}		
		\subfigure[LambdaMART(L).\label{subfig:LambdaMART}]
		{
			\centering{}\includegraphics[width=.24\textwidth, totalheight=1.2in]{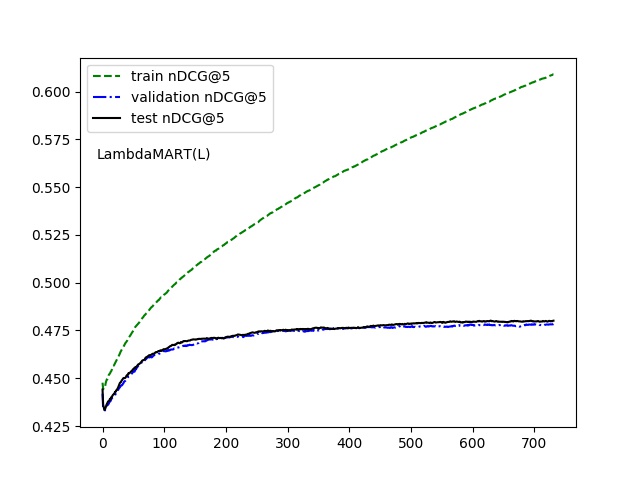}}
		\subfigure[ApproxNDCG.\label{subfig:ApproxNDCG}]
		{
			\centering{}\includegraphics[width=.24\textwidth, totalheight=1.2in]{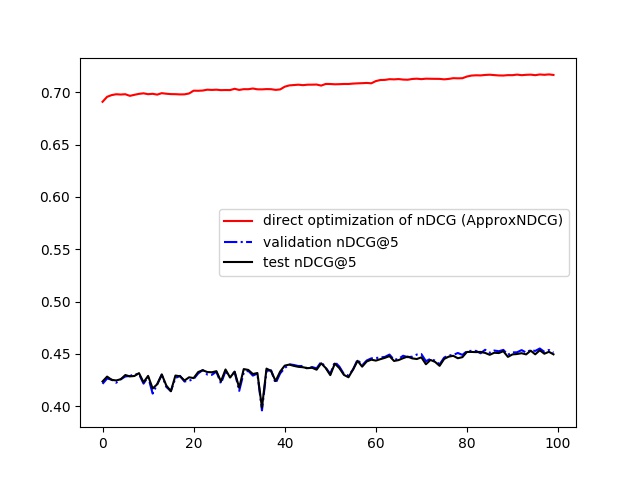}}
		\subfigure[nERR@10-type3.\label{subfig:nERR}]
		{
			\centering{}\includegraphics[width=.24\textwidth, totalheight=1.2in]{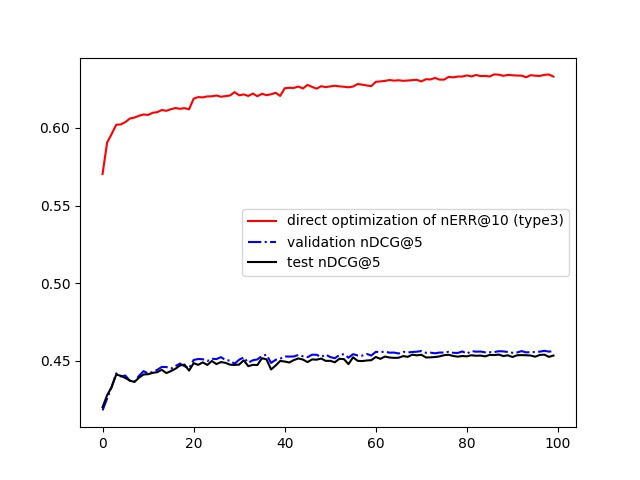}}		
		\subfigure[AP-type3.\label{subfig:AP}]
		{
			\centering{}\includegraphics[width=.24\textwidth, totalheight=1.2in]{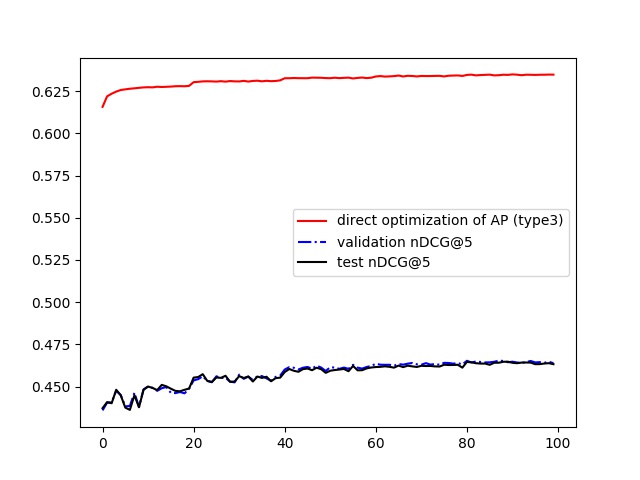}}		
		\subfigure[nDCG-type3.\label{subfig:nDCG}]
		{
			\centering{}\includegraphics[width=.24\textwidth, totalheight=1.2in]{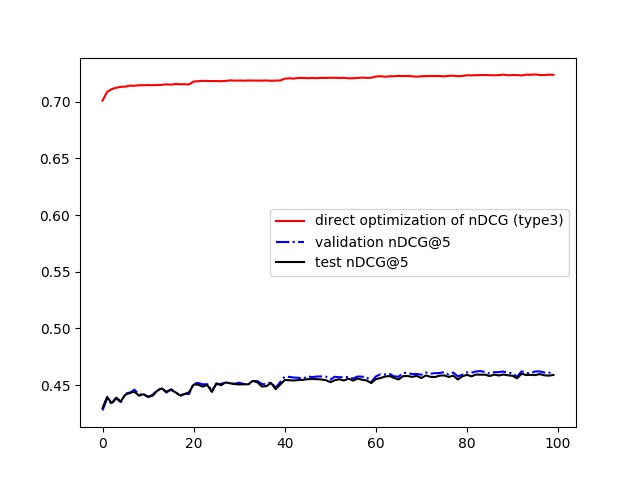}}
		\protect\caption{How the performance (nDCG@5) on the validation and test sets varies as the optimization on the train set progresses. \label{fig:LossCorrelation}}		
		\par\end{centering}
\end{figure}
To well identify the differences between using a surrogate loss and direct metric optimization, it is useful to examine the detailed optimization process. Specifically, we plot how the average loss on the train set varies as optimization progresses, as well as the performance in terms of nDCG@5 on the validation and test sets, where Fold-1 of MSLRWEB30K is used. In Figure \ref{fig:LossCorrelation}, the top row shows the plots of approaches that optimize a surrogate loss, such as WassRank, ListNet, ListMLE. The bottom row shows the plots of methods that directly optimize a specific metric, such as ApproxNDCG, nERR@10-type3, AP-type3 and nDCG-type3. Due to space limitation, not all the aforementioned methods in Section \ref{subsec:OverallCmp} are illustrated. In all plots, the horizontal axis represents the training iteration (epochs for methods based on neural networks and trees for LambdaMART). For ApproxNDCG and the proposed methods that directly optimize a metric, we show the corresponding metric value instead. Following the prior study \cite{RevisitingApproxNDCG}, we plot the performance of LambdaMART(L) in terms of nDCG@5 on the train, validation and test sets, since the gradients of LambdaMART are designed based on some heuristic.

From Figure \ref{fig:LossCorrelation}, we can observe that: (1) Surprisingly, the surrogate loss of WassRank correlates poorly to the evaluation metric, which can not be discovered by merely checking the metric values (e.g., Table \ref{Table:BaselineResults}). In other words, \textit{a rather competitive performance in terms of nDCG does not mean a good match between the surrogate loss and the desired evaluation metric}. For ListNet and ListMLE, the performance in terms of nDCG@5 on the validation and test sets is improved as the surrogate loss decreases. However, \textit{the zigzag shape of the performance curve indicates relatively weak consistency between the surrogate loss and the evaluation metric}. This echoes the finding in prior study \cite{NDCGConsistency}, which proves that ListNet is not consistent with nDCG. LambdaMART(L) shows a sign of overfitting after ensembling a certain number of decision trees (e.g., $300$). (2) For the methods that directly optimize a specific metric, the performance in terms of nDCG@5 on the validation and test sets is improved as the metric value (being optimized) increases. Compared with ApproxNDCG, the performance curves of the proposed methods (nERR@10-type3, AP-type3 and nDCG-type3) are more smooth. Overall, Figure \ref{fig:LossCorrelation} again demonstrates that direct metric optimization is a more appropriate choice than the commonly used surrogate loss functions, such as ListMLE, ListNet and WassRank.
\section{Conclusions}
In this paper, we propose a simple yet effective framework for direct optimization of the widely used IR metrics. The key idea is to use the newly proposed twin-sigmoid function to derive the rank positions of documents during the optimization process. Thanks to the utilization of twin-sigmoid function, on one hand, it enables us to obtain the exact rank position of each document rather than the approximated one. On the other hand, the rank positions derived via twin-sigmoid are differentiable. Then we are able to derive the differentiable reformulations of the widely used IR metrics, such as precision, AP, nERR and nDCG, which can be directly used as the optimization objectives. Furthermore, by carrying out an in-depth analysis of the gradients, we pinpoint the potential limitations inherent with the direct optimization of IR metrics based on vanilla sigmoid. To break the limitations, we propose different strategies to modify the gradient computation. We have shown that the proposed framework leads to substantially improved performance when compared to the previous ranking methods, such as ApproxNDCG, WassRank, ListMLE and ListNet. Compared to the state-of-the-art tree-based approach LambdaMART, the performance of the proposed framework is also comparable. We note that the tree-based model (e.g., LambdaMART) require extensive feature engineering to handle textual features. In contrast, our method building upon neural networks can effectively handle sparse features through embeddings. Also, our analysis indicates that the proposed framework for direction optimization of IR metrics correlates well to the evaluation metric. Since ranking is a core step in a variety of applications, we believe that our framework provides a new perspective for addressing problems of this kind.

For future work, first, we plan to further test the effectiveness of the proposed methods with more datasets across multiple domains, such as Istella LETOR for learning-to-rank and CARS196 for metric learning. Moreover, in terms of neural network design, we did not conduct an in-depth investigation on the impact of different neural architectures due to the high-complex hyperparameter space. From an optimization perspective, there is no guarantee of optimality for a pre-specified architecture like ours in this work. However we do note that the technique of neural architecture search (NAS) \cite{NASSurvey} can be applied. There is some hope that incorporating NAS will make our proposed framework more competitive, which avoids the effort in finding the right network architecture. Second, instead of using neural networks, it is interesting to directly optimize the proposed differentiable reformulations of IR metrics based on gradient boosting decision trees. Then a more fair comparison with LambdaMART(L) can be expected. Third, different from LambdaMART, one potential strength of the proposed framework is the ability of allowing end-to-end direct metric optimization, removing the need for handcrafted features. Therefore, the evaluation of using raw text queries and documents ia also considered as a future work.

\bibliographystyle{unsrt}



\begin{thebibliography}{10}
	
	\bibitem{QuickSort}
	C.~A.~R. Hoare.
	\newblock Quicksort.
	\newblock {\em The {Computer Journal}}, 5(1):10--16, 1962.
	
	\bibitem{l2rbio}
	Bin Liu, Junjie Chen, and Xiaolong Wang.
	\newblock Application of learning to rank to protein remote homology detection.
	\newblock {\em Bioinformatics}, 31(21):3492--3498, 2015.
	
	\bibitem{FstnDCG}
	Kalervo J\"{a}rvelin and Jaana Kek\"{a}l\"{a}inen.
	\newblock Cumulated gain{-}based evaluation of {IR} techniques.
	\newblock {\em ACM Transactions on Information Systems}, 20(4):422--446, 2002.
	
	\bibitem{SubsetRegression}
	David Cossock and Tong Zhang.
	\newblock Subset ranking using regression.
	\newblock In {\em Proceedings of the 19th Annual Conference on Learning
		Theory}, pages 605--619, 2006.
	
	\bibitem{ChuGaussian}
	Wei Chu and Zoubin Ghahramani.
	\newblock Gaussian processes for ordinal regression.
	\newblock {\em Journal of Machine Learning Research}, 6:1019--1041, 2005.
	
	\bibitem{ChuSVOR}
	Wei Chu and {S.}~Sathiya Keerthi.
	\newblock New approaches to support vector ordinal regression.
	\newblock In {\em Proceedings of the 22nd ICML}, pages 145--152, 2005.
	
	\bibitem{FreundBoosting}
	Yoav Freund, Raj Iyer, Robert~E. Schapire, and Yoram Singer.
	\newblock An efficient boosting algorithm for combining preferences.
	\newblock {\em Journal of Machine Learning Research}, 4:933--969, 2003.
	
	\bibitem{ShenPerceptron}
	Libin Shen and Aravind~K. Joshi.
	\newblock Ranking and reranking with perceptron.
	\newblock {\em Machine Learning}, 60(1-3):73--96, 2005.
	
	\bibitem{RankSVMStruct}
	Thorsten Joachims.
	\newblock Training linear {SVMs} in linear time.
	\newblock In {\em Proceedings of the 12th KDD}, pages 217--226, 2006.
	
	\bibitem{OlivierMargin}
	Olivier Chapelle, Quoc Le, and Alex Smola.
	\newblock Large margin optimization of ranking measures.
	\newblock In {\em NIPS workshop on Machine Learning for Web Search}, 2007.
	
	\bibitem{AdaRank}
	Jun Xu and Hang Li.
	\newblock Adarank: a boosting algorithm for information retrieval.
	\newblock In {\em Proceedings of the 30th SIGIR}, pages 391--398, 2007.
	
	\bibitem{YueSVMAP}
	Yisong Yue, Thomas Finley, Filip Radlinski, and Thorsten Joachims.
	\newblock A support vector method for optimizing average precision.
	\newblock In {\em Proceedings of the 30th SIGIR}, pages 271--278, 2007.
	
	\bibitem{GuiverSoftRank}
	John Guiver and Edward Snelson.
	\newblock Learning to rank with softrank and gaussian processes.
	\newblock In {\em Proceedings of the 31st SIGIR}, pages 259--266, 2008.
	
	\bibitem{SoftRank}
	Michael Taylor, John Guiver, Stephen Robertson, and Tom Minka.
	\newblock Softrank{:} optimizing non-smooth rank metrics.
	\newblock In {\em Proceedings of the 1st WSDM}, pages 77--86, 2008.
	
	\bibitem{QinApproximateNDCG}
	Tao Qin, Tie{-}Yan Liu, and Hang Li.
	\newblock A general approximation framework for direct optimization of
	information retrieval measures.
	\newblock {\em Journal of Information Retrieval}, 13(4):375--397, 2010.
	
	\bibitem{LambdaMART}
	Qiang Wu, Christopher~J. Burges, Krysta~M. Svore, and Jianfeng Gao.
	\newblock Adapting boosting for information retrieval measures.
	\newblock {\em Journal of Information Retrieval}, 13(3):254--270, 2010.
	
	\bibitem{ListNet}
	Zhe Cao, Tao Qin, Tie{-}Yan Liu, Ming{-}Feng Tsai, and Hang Li.
	\newblock Learning to rank{:} from pairwise approach to listwise approach.
	\newblock In {\em Proceedings of the 24th ICML}, pages 129--136, 2007.
	
	\bibitem{ListMLE}
	Fen Xia, Tie{-}Yan Liu, Jue Wang, Wensheng Zhang, and Hang Li.
	\newblock Listwise approach to learning to rank{:} theory and algorithm.
	\newblock In {\em Proceedings of the 25th ICML}, pages 1192--1199, 2008.
	
	\bibitem{BoltzRank}
	Maksims~N{.} Volkovs and Richard~S{.} Zemel.
	\newblock Boltzrank{:} learning to maximize expected ranking gain.
	\newblock In {\em Proceedings of ICML}, pages 1089--1096, 2009.
	
	\bibitem{LambdaRank}
	Christopher~{J.C.} Burges, Robert Ragno, and Quoc~Viet Le.
	\newblock Learning to rank with nonsmooth cost functions.
	\newblock In {\em Proceedings of NeurIPS}, pages 193--200, 2006.
	
	\bibitem{ListwiseNeuralDemo}
	Razieh Rahimi, Ali Montazeralghaem, and James Allan.
	\newblock Listwise neural ranking models.
	\newblock In {\em Proceedings of ICTIR 2019}, pages 101--104, 2019.
	
	\bibitem{TaoWSDM2019}
	Hai{-}Tao Yu, Adam Jatowt, Hideo Joho, Joemon Jose, Xiao Yang, and Long Chen.
	\newblock Wassrank{:} listwise document ranking using optimal transport theory.
	\newblock In {\em Proceedings of the 12th WSDM}, pages 24--32, 2019.
	
	\bibitem{OlivierSmoothedIRMetric}
	Olivier Chapelle and Mingrui Wu.
	\newblock Gradient descent optimization of smoothed information retrieval
	metrics.
	\newblock {\em Journal of Information Retrieval}, 13(3):216--235, 2010.
	
	\bibitem{LMarginORM}
	Olivier Chapelle, Quoc Le, and Alex Smola.
	\newblock Large margin optimization of ranking measures.
	\newblock In {\em NIPS 2007 Workshop on Machine Learning for Web Search}, 2007.
	
	\bibitem{DirOptRM}
	Quoc Le and Alex Smola.
	\newblock Direct optimization of ranking measures.
	\newblock {\em arXiv:0704.3359v1}, 2007.
	
	\bibitem{OptRM}
	Quoc Le, Alex Smola, Olivier Chapelle, and Choon~Hui Teo.
	\newblock Optimization of ranking measures.
	\newblock {\em Journal of Machine Learning Research}, pages 1--48, 2010.
	
	\bibitem{SmoothHinge}
	Mingrui Wu, Yi~Chang, Zhaohui Zheng, and Hongyuan Zha.
	\newblock Smoothing {DCG} for learning to rank{:} a novel approach using
	smoothed hinge functions.
	\newblock In {\em Proceedings of the 18th CIKM}, pages 1923--1926, 2009.
	
	\bibitem{AppBridges}
	Jason Ramapuram and Russ Webb.
	\newblock Differentiable approximation bridges for training networks containing
	non{-}differentiable functions.
	\newblock {\em arXiv:1905.03658}, 2019.
	
	\bibitem{OptBlackboxMetric}
	Qijia Jiang, Olaoluwa Adigun, Harikrishna Narasimhan, Mahdi~Milani Fard, and
	Maya Gupta.
	\newblock Optimizing black{-}box metrics with adaptive surrogates.
	\newblock {\em arXiv:2002.08605}, 2020.
	
	\bibitem{L2LGD}
	Marcin Andrychowicz, Misha Denil, Sergio G\'{o}mez, Matthew~W Hoffman, David
	Pfau, Tom Schaul, Brendan Shillingford, and Nando de~Freitas.
	\newblock Learning to learn by gradient descent by gradient descent.
	\newblock In {\em Proceedings of NeurIPS}, pages 3981--3989, 2016.
	
	\bibitem{LambdaLossFramework}
	Xuanhui Wang, Cheng Li, Nadav Golbandi, Michael Bendersky, and Marc Najork.
	\newblock The lambdaloss framework for ranking metric optimization.
	\newblock In {\em Proceedings of the 27th CIKM}, pages 1313--1322, 2018.
	
	\bibitem{XuDirOptMetrics}
	Jun Xu, Tie{-}Yan Liu, Min Lu, Hang Li, and Wei{-}Ying Ma.
	\newblock Directly optimizing evaluation measures in learning to rank.
	\newblock In {\em Proceedings of SIGIR}, pages 107--114, 2008.
	
	\bibitem{KDDOptRM}
	Ming Tan, Tian Xia, Lily Guo, and Shaojun Wang.
	\newblock Direct optimization of ranking measures for learning to rank models.
	\newblock In {\em Proceedings of the 19th KDD}, pages 856--864, 2013.
	
	\bibitem{HEClonal}
	Qiang He, Jun Ma, and Shuaiqiang Wang.
	\newblock Directly optimizing evaluation measures in learning to rank based on
	the clonal selection algorithm.
	\newblock In {\em Proceedings of the 19th CIKM}, pages 1449--1452, 2010.
	
	\bibitem{NDCGTiedScores}
	Andrey Kustarev, Yury Ustinovsky, Yury Logachev, Evgeny Grechnikov, Ilya
	Segalovich, and Pavel Serdyukov.
	\newblock Smoothing {NDCG} metrics using tied scores.
	\newblock In {\em Proceedings of CIKM}, pages 2053--2056, 2011.
	
	\bibitem{HamedL2rMetric}
	Hamed Valizadegan, Rong Jin, Ruofei Zhang, and Jianchang Mao.
	\newblock Learning to rank by optimizing {NDCG} measure.
	\newblock In {\em Proceedings of 22nd NIPS}, pages 1883--1891, 2009.
	
	\bibitem{RevisitingApproxNDCG}
	Sebastian Bruch, Masrour Zoghi, Michael Bendersky, and Marc Najork.
	\newblock Revisiting approximate metric optimization in the age of deep neural
	networks.
	\newblock In {\em Proceedings of the 42nd SIGIR}, pages 1241--1244, 2019.
	
	\bibitem{DSSM}
	Po-Sen Huang, Xiaodong He, Jianfeng Gao, Li~Deng, Alex Acero, and Larry Heck.
	\newblock Learning deep structured semantic models for web search using
	clickthrough data.
	\newblock In {\em Proceedings of CIKM}, pages 2333--2338, 2013.
	
	\bibitem{CDSSM}
	Yelong Shen, Xiaodong He, Jianfeng Gao, Li~Deng, and Gr{\'e}goire Mesnil.
	\newblock Learning semantic representations using convolutional neural networks
	for web search.
	\newblock In {\em Proceedings of the 23rd WWW}, pages 373--374, 2014.
	
	\bibitem{DRMM}
	Jiafeng Guo, Yixing Fan, Qingyao Ai, and W.~Bruce Croft.
	\newblock A deep relevance matching model for {Ad-hoc} retrieval.
	\newblock In {\em Proceedings of the 25th CIKM}, pages 55--64, 2016.
	
	\bibitem{HuCNNMatching}
	Baotian Hu, Zhengdong Lu, Hang Li, and Qingcai Chen.
	\newblock Convolutional neural network architectures for matching natural
	language sentences.
	\newblock In {\em Proceedings of 27th NIPS}, pages 2042--2050, 2014.
	
	\bibitem{PangAAAMatching}
	Liang Pang, Yanyan Lan, Jiafeng Guo, Jun Xu, Shengxian Wan, and Xueqi Cheng.
	\newblock Text matching as image recognition.
	\newblock In {\em Proceedings of AAAI Conference on Artificial Intelligence},
	pages 2793--2799, 2016.
	
	\bibitem{MatchSRNN}
	Shengxian Wan, Yanyan Lan, Jun Xu, Jiafeng Guo, Liang Pang, and Xueqi Cheng.
	\newblock Match{-}srnn{:} modeling the recursive matching structure with
	spatial rnn.
	\newblock In {\em Proceedings of IJCAI conference}, pages 2922--2928, 2016.
	
	\bibitem{Seq2Slate}
	Irwan Bello, Sayali Kulkarni, Sagar Jain, Craig Boutilier, Ed~Chi, Elad Eban,
	Xiyang Luo, Alan Mackey, and Ofer Meshi.
	\newblock Seq2slate{:} re{-}ranking and slate optimization with {RNNs}.
	\newblock In {\em Proceedings of the Workshop on Negative Dependence in Machine
		Learning}, 2019.
	
	\bibitem{IRJNeuralIR}
	Kezban~Dilek Onal, Ye~Zhang, Ismail~Sengor Altingovde, et~al.
	\newblock Neural information retrieval{:} at the end of the early years.
	\newblock {\em Journal of Information Retrieval}, 21(2-3):111--182, 2018.
	
	\bibitem{IPMNeuralIR}
	Jiafeng Guo, Yixing Fan, Liang Pang, Liu Yang, Qingyao Ai, Hamed Zamani, Chen
	Wu, W.~Bruce Croft, and Xueqi Cheng.
	\newblock A deep look into neural ranking models for information retrieval.
	\newblock {\em Information Processing {\&} Management}, 2019.
	
	\bibitem{LinBERTDocR}
	Zeynep~Akkalyoncu Yilmaz, Shengjin Wang, Wei Yang, Haotian Zhang, and Jimmy
	Lin.
	\newblock Applying {BERT} to document retrieval with birch.
	\newblock In {\em Proceedings of {EMNLP} 2019}, pages 19--24, 2019.
	
	\bibitem{RerankingBERT}
	Rodrigo Nogueira and Kyunghyun Cho.
	\newblock Passage re-ranking with {BERT}.
	\newblock {\em arXiv:1901.04085v4}, 2019.
	
	\bibitem{CEDR}
	Sean MacAvaney, Andrew Yates, Arman Cohan, and Nazli Goharian.
	\newblock {CEDR:} contextualized embeddings for document ranking.
	\newblock In {\em Proceedings of the 42nd SIGIR}, pages 1101--1104, 2019.
	
	\bibitem{BERT}
	Jacob Devlin, Ming{-}Wei Chang, Kenton Lee, and Kristina Toutanova.
	\newblock {BERT}{:} pre{-}training of deep bidirectional transformers for
	language understanding.
	\newblock In {\em Proceedings of {NAACL-HLT} 2019}, pages 4171--4186, 2019.
	
	\bibitem{NDCGConsistency}
	Pradeep Ravikumar, Ambuj Tewari, and Eunho Yang.
	\newblock On {NDCG} consistency of listwise ranking methods.
	\newblock In {\em Proceedings of Machine Learning Research}, number~15, pages
	618--626, 2011.
	
	\bibitem{ERR}
	Olivier Chapelle, Donald Metlzer, Ya~Zhang, and Pierre Grinspan.
	\newblock Expected reciprocal rank for graded relevance.
	\newblock In {\em Proceedings of the 18th CIKM}, pages 621--630, 2009.
	
	\bibitem{EmpiricalSmoothMetric}
	Yisong Yue and Christopher~{J.C.} Burges.
	\newblock On using simultaneous perturbation stochastic approximation for
	learning to rank, and the empirical optimality of lambdarank.
	\newblock In {\em Microsoft Research Technical Report MSR-TR-2007-115}, 2007.
	
	\bibitem{LETORIR}
	Tao Qin, Tie-Yan Liu, Jun Xu, and Hang Li.
	\newblock {LETOR:} a benchmark collection for research on learning to rank for
	information retrieval.
	\newblock {\em Information Retrieval Journal}, 13(4):346--374, 2010.
	
	\bibitem{StochasticTreatmentRF}
	Sebastian Bruch, Shuguang Han, Michael Bendersky, and Marc Najork.
	\newblock A stochastic treatment of learning to rank scoring functions.
	\newblock In {\em Proceedings of the 13th WSDM}, pages 61--69, 2020.
	
	\bibitem{LightGBM}
	Guolin Ke, Qi~Meng, Thomas Finley, Taifeng Wang, Wei Chen, Weidong Ma, Qiwei
	Ye, and Tie{-}Yan Liu.
	\newblock Lightgbm{:} {A} highly efficient gradient boosting decision tree.
	\newblock In {\em Proceedings of NeurIPS}, pages 3149--3157, 2017.
	
	\bibitem{PListMLE}
	Yanyan Lan, Yadong Zhu, Jiafeng Guo, Shuzi Niu, and Xueqi Cheng.
	\newblock Position{-aware ListMLE:} a sequential learning process for ranking.
	\newblock In {\em Proceedings of the 30th Conference on UAI}, pages 449--458,
	2014.
	
	\bibitem{NASSurvey}
	Thomas Elsken, Jan~Hendrik Metzen, and Frank Hutter.
	\newblock Neural architecture search{:} a survey.
	\newblock {\em Journal of Machine Learning Research}, 20(55):1--21, 2019.
	
\end{thebibliography}

\appendix

\section{Sigmoid and The Proposed Twin-Sigmoid}
The traditional \textit{sigmoid} function is written as follows:
\begin{equation}
\sigma(z,\alpha)=\frac{1}{1+\exp(-\alpha\cdot z)}
\end{equation}
\noindent where the constant $\alpha>0$ controls how steep is the sigmoid. Its derivative is computed as:
\begin{equation}
\frac{\partial\sigma(z,\alpha)}{\partial z}=\alpha\cdot\sigma(z,\alpha)\cdot[1-\sigma(z,\alpha)]
\end{equation}

The proposed \textit{twin-sigmoid} function (referred to as $\sigma^{+}$) consists of two ordinary sigmoid functions. The \textit{forward sigmoid} function (referred to as $\sigma^{+}_{f}$) with a sufficient large scalar (denoted as $\alpha_{f}$) is responsible for generating the output, namely
\begin{equation}
\sigma^{+}(z,\alpha_{f},\alpha_{b})=\sigma_{f}^{+}(z,\alpha_{f})=\frac{1}{1+\exp(-\alpha_{f}\cdot z)}
\end{equation} 

The \textit{backward sigmoid} function (referred to as $\sigma^{+}_{b}$ ) with a small scalar (denoted as $\alpha_{b}$) is responsible for generating the gradient for back-propagation, namely
\begin{equation}
\sigma_{b}^{+}(z,\alpha_{b})=\frac{1}{1+\exp(-\alpha_{b}\cdot z)}
\end{equation}
\begin{equation}
\frac{\partial\sigma^{+}(z,\alpha_{f},\alpha_{b})}{\partial z}=\alpha_{b}\cdot\sigma_{b}^{+}(z,\alpha_{b})\cdot[1-\sigma_{b}^{+}(z,\alpha_{b})]
\end{equation}

\section{Gradients}
In this section, we explain in full detail how the gradients are computed when optimizing the smoothed IR metrics based on the stochastic gradient descent algorithm.

Given the differentiable formulation of predicted rank position $r_{i}=1+\sum_{j:j\neq i}1-\sigma^{+}(y_{ij})$ for the document $x_{i}$, we have
\begin{equation}
\frac{\partial r_{i}}{\partial y_{i}}=\sum_{j:j\neq i}-\alpha_{b}\cdot \sigma^{+}_{b}(y_{ij})\cdot[1-\sigma^{+}_{b}(y_{ij})]
\end{equation}
\begin{equation}
\frac{\partial r_{i}}{\partial y_{j}}=\alpha_{b}\cdot \sigma^{+}_{b}(y_{ij})\cdot[1-\sigma^{+}_{b}(y_{ij})]
\end{equation}
where $\sigma^{+}_{b}$ denotes the backward sigmoid component of $\sigma^{+}$.

For the metric of precision, the gradient with respect to the rank position is 

\begin{equation}
\frac{\partial\widehat{Pre}@k}{\partial\bar{r_{i}}}=\frac{b_{i}^{**}\cdot i}{k}\cdot\frac{-1}{(\bar{r_{i}})^{2}}=-\frac{b_{i}^{**}}{k\cdot\bar{r_{i}}}
\end{equation}

For the metric of AP, Eq-\ref{Eq:VirAP} can be rewritten as

\begin{equation}
\widehat{AP}=\frac{1}{|Y^{+}|}\sum_{i=1}^{m}b_{i}^{**}\cdot\frac{i}{\bar{r_{i}}}(\sum_{j=i}^{m}\frac{b_{j}^{**}}{j})
\end{equation}

then we compute the gradient with respect to the rank position as

\begin{equation}
\frac{\partial\widehat{AP}}{\partial\bar{r_{i}}}=\frac{b_{i}^{**}\cdot i\cdot\sum_{j=i}^{m}\frac{b_{j}^{**}}{j}}{|Y^{+}|}\cdot\frac{-1}{(\bar{r_{i}})^{2}}=-\frac{b_{i}^{**}\cdot\sum_{j=i}^{m}\frac{b_{j}^{**}}{j}}{|Y^{+}|}\cdot\frac{1}{\bar{r_{i}}}
\end{equation}

For the metric of nDCG, the gradient with respect to the rank position is computed as

\begin{equation}
\frac{\partial\widehat{nDCG}}{\partial r_{k}}=\frac{2^{y_{k}^{*}}-1}{DCG^{*}}\cdot\frac{-1}{(\log_{2}(r_{k}+1))^{2}}\cdot\frac{1}{r_{k}+1}
\end{equation}

For the metric of nERR, the gradient with respect to the rank position is

\begin{equation}
\frac{\partial\widehat{nERR}@k}{\partial\bar{r_{j}}}=\frac{1}{ERR^{*}@k}\cdot\prod_{i=1}^{j-1}(1-\frac{2^{y_{i}^{**}}-1}{2^{\max(\mathbf{y}^{**})}})\cdot\frac{2^{y_{j}^{**}}-1}{2^{\max(\mathbf{y}^{**})}}\cdot\frac{-1}{(\bar{r_{j}})^{2}}=-\frac{\prod_{i=1}^{j-1}(1-\frac{2^{y_{i}^{**}}-1}{2^{\max(\mathbf{y}^{**})}})\cdot\frac{2^{y_{j}^{**}}-1}{2^{\max(\mathbf{y}^{**})}}}{ERR^{*}@k}\cdot\frac{1}{(\bar{r_{j}})^{2}}
\end{equation}

By viewing the \textit{negative metric score} as the ranking loss, according to the chain rule, it is straightforward to compute the gradients of the ranking loss.
\end{document}